\documentclass[prd,aps,showpacs,epsf,floats,onecolumn]{revtex4}%
\usepackage{amssymb}
\usepackage{amsfonts}
\usepackage{amsmath}
\usepackage{graphicx}%
\providecommand{\U}[1]{\protect\rule{.1in}{.1in}}
\begin{document}
\title{\textbf{Violation of Bell's Inequality in the Clauser-Horne-Shimony-Holt Form
with Entangled Quantum States Revisited}}
\author{\textbf{Carlo Cafaro}$^{1,2}$, \textbf{Christian Corda}$^{2,3,4}$,
\textbf{Philip Cairns}$^{5}$, \textbf{Ayhan Bingolbali}$^{6,1}$}
\affiliation{$^{1}$University at Albany-SUNY, 12222 Albany, New York, USA}
\affiliation{$^{2}$SUNY Polytechnic Institute, 13502 Utica, New York, USA}
\affiliation{$^{3}$Istituto Livi, 59100 Prato, Prato, Italy}
\affiliation{$^{4}$International Institute for Applicable Mathematics and Information
Sciences, 500063 Adarsh Nagar, Hyderabad, India}
\affiliation{$^{5}$Physics Program, Graduate Center of the City University of New York, New
York, NY 10016, USA}
\affiliation{$^{6}$Yildiz Technical University, 34220 Istanbul, Turkey}

\begin{abstract}
Scientific imagination and experimental ingenuity are at the heart of physics.
One of the most known instances where this interplay between theory (i.e.,
foundations) and experiments (i.e., technology) occurs is in the discussion of
Bell's inequalities. In this paper, we present a revisitation of the violation
of Bell's inequality in the Clauser-Horne-Shimony-Holt (CHSH) form with
entangled quantum states. First, we begin with a discussion of the 1935
Einstein-Podolski-Rosen (EPR) paradox (i.e., incompleteness of quantum
mechanics) that emerges from putting the emphasis on Einstein's locality and
the absolute character of physical phenomena. Second, we discuss Bell's 1971
derivation of the 1969 CHSH form of the original 1964 Bell inequality in the
context of a realistic local hidden-variable theory (RLHVT). Third,
identifying the quantum-mechanical spin correlation coefficient with the RLHVT
one, we follow Gisin's 1991 analysis to show that quantum mechanics violates
Bell's inequality when systems are in entangled quantum states. For
pedagogical purposes, we show how the extent of this violation depends both on
the orientation of the polarizers and the degree of entanglement of the
quantum states. Fourth, we discuss the basics of the experimental verification
of Bell's inequality in an actual laboratory as presented in the original 1982
Aspect-Grangier-Roger (AGR) experiment. Finally, we provide an outline of some
essential take home messages from this wonderful example of physics at its best.

\end{abstract}

\maketitle

\section{Introduction}

Information theory, quantum theory, and relativity theory are essential
ingredients of theoretical physics \cite{peres2004}. Theoretical physics aims
at describing and, to a certain extent, comprehending natural events.
Unfortunately, issues emerge when one tries to combine general relativity (GR)
theory with quantum theory, two main pillars of modern physics. During the
last fifty years, quantum theory and general relativity have been the object
of an important scientific discussion by theoretical physicists \cite{Corda2}.
For details on a general discussion on the quantum gravity problem, see
\cite{Corda2}. Here we limit ourselves to release some general considerations.
On the one hand, general relativity well-describes gravitational phenomena at
large scales, starting from observations from cosmological distances to
millimeter scales. On the other hand, quantum mechanics and quantum field
theory well-specify phenomena at small scales from a fraction of a millimeter
down to $10^{-19}$ meters. Such scales are dominated by strong and electroweak
interactions. From a historical point of view, general relativity obtained
great success. For instance, the famous Soviet physicist Lev Davidovich Landau
used to say that Einstein's theory of gravity was the world's best scientific
theory on a par with quantum mechanics \cite{Landau}. While GR has correctly
predicted theoretical results that are consistent with experiments and
observations, it has shown a variety of limitations and weaknesses as well.
Therefore, gravitational physicists can question whether nowadays GR has a
definitive behavior \cite{Corda}. Unlike other theories (such as, for
instance, electromagnetism) general relativity has not been quantized so far.
This issue does not allow to treat gravity like other quantized theories.
This, in turn, constitutes a strong obstacle for the path leading to the
unification of gravitation with other forces. So far, none has been able to
realize a consistent quantum gravity theory capable of leading to the
unification of gravity with the other fundamental interactions. A key point to
understand seems to be the role played by black holes in a quantum gravity
setting. Indeed, researchers in quantum gravity have nowadays the general
conviction that black holes should be the fundamental bricks of quantum
gravity, just like atoms have been the fundamental bricks of quantum mechanics
at the outset of quantum theory. This conviction originated from an original
intuition of Bekenstein \cite{Bekenstein}.

Historically and while seeking a unifying theory of all fundamental
interactions, Einstein had the conviction that quantum mechanics could not be
considered as being complete. His opinion was that scientists should attempt
to \textquotedblleft test\textquotedblright\ quantum mechanics via a final
deterministic theory which he labelled \emph{Generalized Theory of
Gravitation} as pointed out in Ref. \cite{Pais}. Indeed, Einstein attempted to
follow such a path. However, he failed in obtaining the final equations of
such a unified field theory \cite{Pais}. Nowadays, Einstein's opinion is still
partially endorsed by some scientists \cite{Corda2}. The problem of a missing
quantization of gravity is present and\textbf{ }obstacles emerge when one
tries to combine GR with quantum theory. For instance, in classical Newtonian
gravity the mass does not enter Newton's equation of motion,%
\begin{equation}
\frac{d^{2}\vec{x}}{dt^{2}}=\vec{g}\text{,} \label{m1}%
\end{equation}
with $\vec{g}$ denoting the (constant) acceleration of gravity and $\vec
{x}=\vec{x}\left(  t\right)  $ being the particle's position vector. The lack
of the mass term in Eq. (\ref{m1}) is justified by the equality of the
gravitational and inertial masses (Equivalence Principle). The scenario is
different in quantum mechanics where, focusing on a particle in an external
gravitational potential field $\Phi_{\text{gravity}}$, the mass $m$ no longer
cancels in Schr\"{o}dinger's quantum-mechanical wave equation \cite{sakurai},%
\begin{equation}
\left[  -\frac{\hbar^{2}}{2m}\vec{\nabla}_{\vec{x}}^{2}+m\Phi_{\text{gravity}%
}\right]  \psi\left(  \vec{x}\text{, }t\right)  =i\hbar\frac{\partial
}{\partial t}\psi\left(  \vec{x}\text{, }t\right)  \text{,} \label{m2}%
\end{equation}
with $\psi=\psi\left(  \vec{x}\text{, }t\right)  $ being the particle's wave
function. Unlike Eq. (\ref{m1}), the mass $m$ does not disappear in Eq.
(\ref{m2}). It manifests itself in the combination $\hbar/m$, where
$\hbar\overset{\text{def}}{=}h/2\pi$ and $h$ denotes the Planck constant. From
Eq. (\ref{m2}), is seems that $m$ is expected to appear when $\hbar$ appears.
It seems reasonable to think that the challenges in merging a classical theory
of gravity and quantum theory in a single unified theory are not simply
technical and mathematical, but rather conceptual and fundamental
\cite{colella1975}. In fact, from the above example one could expect a
breakdown of the Equivalence Principle in the quantum framework, despite there
is no experimental evidence in that sense, because the Equivalence Principle
is today tested with a very high precision \cite{Corda2, MTW}. In recent
years, it was proposed in Ref. \cite{brukner14} that \emph{quantum causality}
might help shedding some light on foundational issues emerging from the
general relativity-quantum mechanics combination problem. This is in agreement
with the above cited point of view of Einstein \cite{Pais}.

When characterizing physical phenomena at the quantum level, the interaction
between the mechanical object under study and the observer (or, alternatively,
observer's measuring equipment) is not negligible and cannot be predicted
\cite{bohr35, bohr37, bohr50}. The presence of such a nonnegligible
interaction yields the impossibility of unambiguously discriminating between
the object and the measuring instruments. However, the classical concept of
causality demands the possibility of sharply distinguishing between the
subject and the object. Therefore, the above mentioned impossibility of
unambiguous discrimination in a quantum-mechanical setting is not logically
compatible with the classical notion of causality. While trying to bring
consistency in science, Bohr proposed to substitute the classical ideal of
causality with a more general concept termed \emph{complementarity}. In rather
simple words, it is clear that an individual cannot bow in front of somebody
without showing one's back to somebody else. This simplifying statement is at
the roots of one of the most revolutionary scientific concepts of the
twentieth century, namely Bohr's complementarity principle \cite{wheeler63}.
More specifically, this principle encodes an essential feature of quantum
physics and represents the dichotomy between the corpuscular (i.e., particle)
and ondulatory (i.e., wave) nature of quantum-mechanical objects (both matter
and light). Within this dichotomy descriptive setting, particle and wave
properties are specified by well-defined position and momentum, respectively
\cite{vaccaro10}. In the famous 1935 \emph{EPR paradox} \cite{epr}, Einstein,
Podolsky, and Rosen arrived at the conclusion that the quantum-mechanical
description of physical reality provided by wave functions was incomplete
based on a line of reasoning that relied on a gedankenexperiment (i.e.,
thought experiment). Then, employing the complementarity principle along with
providing a different interpretation of the notion of locality \cite{bohr35},
Bohr criticized the EPR argumentation.

It is worth emphasizing that it is possible to formulate many different
versions of the original 1935 EPR paradox \cite{omnes}. Given its historical
relevance, we would like to briefly mention Bohm's 1951 version of the EPR
paradox \cite{Bohm,bohm57,Laloe}. In Bohm's version of the paradox, pairs of
discrete variables (such as, spin components $\left\{  s_{i}\right\}  _{1\leq
i\leq3}$ along distinct directions with $\left[  s_{i}\text{, }s_{j}\right]
=i\hslash\varepsilon_{ijk}s_{k}$) replace pairs of continuous variables (such
as, position $\left\{  x_{i}\right\}  _{1\leq i\leq3}$ and momentum $\left\{
p_{i}\right\}  _{1\leq i\leq3}$ with $\left[  x_{i}\text{, }p_{j}\right]
=i\hslash\delta_{ij}$). Bohm considered the decay of a spin-$0$ particle into
a pair of spin-$1/2$ particles, such as the decay of a neutral pion $\pi^{0}$
into a positron-electron pair (i.e., $\pi^{0}\rightarrow e^{+}e^{-}$). As a
preliminary remark, we recall that quantum mechanics requires that at most one
spin component of each particle may be definite and be predictable with no
uncertainty, since spin components along distinct directions constitute
incompatible observables. Assume that after the decay the two particles (i.e.,
$e^{+}$ and $e^{-}$) are very far apart. Assume also to measure the
$x$-component $s_{x}^{e^{-}}$ of the electron and find $s_{x}^{e^{-}}%
=+\hslash/2$. Then, since the total spin is assumed to be conserved and since
the measurement on $e^{-}$ is supposed to not affect the real factual
situation of $e^{+}$, one knows with certainty that $s_{x}^{e^{+}}=-\hslash
/2$. Additionally, one could have measured $s_{y}^{e^{-}}$ and, using the same
line of reasoning, $s_{y}^{e^{+}}$ would have been definite; and similarly for
$s_{z}^{e^{+}}$. This scenario is paradoxical since it seems that one can
predict with no uncertainty the values of all three spin components
$s_{x}^{e^{+}}$, $s_{y}^{e^{+}}$, and $s_{z}^{e^{+}}$, despite the fact that
they are non-commuting observables. We shall see that the solution to this
inconsistency lies in an unsound line of reasoning. First, one is not taking
into account the fact the there are no quantum measurement outcomes without
actually performing an experiment \cite{peres78}. Second, one is neglecting
the fact that the choice of the experiment performed on $e^{-}$ dictates the
prediction that can be made for the observations of the experiment performed
on $e^{+}$, being the composite quantum system $\left(  e^{-}\text{, }%
e^{+}\right)  $ entangled. Interestingly, quantum entanglement seems to play a
key role in a possible solution to Hawking's information paradox in the
context of black holes physics \cite{Hawking}. Indeed, the original paradox
relies on the (faulty) factorization assumption of the radiation. However, as
reported in Ref. \cite{calmet22}, black hole information is encoded in
entangled macroscopic superposition states of the radiation.

In 1964, John Bell solved the EPR paradox in Ref. \cite{bell64}. The concept
of causality played a crucial role in the EPR paradox, Bohr's complementarity
principle and, finally, Bell's theorem.

The literature on Bell's inequality is vast and we shall not try to cite it
all here. However, as students, teachers, and researchers, we really enjoyed
the presentation of the 1991 Gisin work in Ref. \cite{gisin91}. For this
reason, we presented our first mathematical reconsideration of Gisin's work in
Ref. \cite{cafaro16}. There, we reexamined Gisin's 1991 proof regarding the
violation of Bell's inequality for any pure entangled state of two-particle
systems. Our investigation was motivated by didactic reasons and permitted to
straighten a few mathematical points in the original proof that in no way
modified the physical content provided by Gisin's work. In recent years
\cite{elena}, it was brought to our attention that in both Gisin's 1991 work
and our 2016 work, there was an incorrect mathematical inequality that
remained to be fixed. The work that we present in this paper is mainly
motivated by our desire of fixing this mathematical imperfection along with
the will of providing a better physical interpretation of our own work in Ref.
\cite{cafaro16}. In short, our goals can be summarized as follows:

\begin{enumerate}
\item[{[i]}] Present \ a correct mathematical treatment of Gisin's 1991 proof
of Bell's inequality following the works in Refs. \cite{gisin91,cafaro16}.

\item[{[ii]}] Enhance the self-sufficiency of the presentation in Refs.
\cite{gisin91,cafaro16} by providing a derivation of Bell's inequality in the
so-called Clauser-Horne-Shimony-Holt (CHSH) form \cite{clauser69}, the one
originally used by Gisin in his 1991 work \cite{gisin91}.

\item[{[iii]}] Clarify, within a unifying setting with both conceptual and
experimental insights, the comparison between theoretical predictions that
emerge from quantum mechanics (QM) and a realistic hidden variable theory
(RLHVT) with actual experimental observations carried out in the well-known
1982 Aspect-Grangier-Roger (AGR) experiment \cite{aspect82}.
\end{enumerate}

The rest of this paper is organized as follows. In Section II, we discuss the
1935 Einstein-Podolski-Rosen (EPR) paradox. In Section III, we revisit Bell's
1971 derivation of the 1969 Clauser-Horne-Shimony-Holt (CHSH) form of the
original 1964 Bell inequality in the framework of realistic local
hidden-variable theories. In Section IV, after identifying the
quantum-mechanical spin correlation coefficient with the RLHVT one and, in
addition, following the path traced by Gisin's 1991 analysis \cite{gisin91},
we verify in an explicit manner that quantum mechanics violates Bell's
inequality when systems are in entangled quantum states. Furthermore, we
illustrate with a simple set of examples how the extent of this Bell violation
depends both on the orientation of the polarizers and the entanglement degree
of the quantum states specifying the physical systems being considered. In
Section V, we present the essential features of the experimental verification
of Bell's inequality as originally proposed by Aspect-Grangier-Roger in their
1982 experiment. Finally, we provide in Section VI an outline of some
essential take home messages that originate from this wonderful example of
physics at its best.

\section{The EPR paradox}

\begin{table}[t]
\centering
\begin{tabular}
[c]{c|c}\hline\hline
\textbf{Property} & \textbf{Meaning}\\\hline
Realism & Systems have local objective properties, independent of
observation\\
Einstein locality & Superluminal communication is impossible\\
Bell locality & Factorization of the joint probability distribution of the
properties of a composite system\\
Quantum nonlocality & Entanglement allows instantaneous propagation of
correlations among properties\\\hline
\end{tabular}
\caption{Schematic description of the meaning of key concepts that appear in
the Bell inequality discussion: Realism, Einstein locality, Bell locality and,
finally, quantum nonlocality.}%
\end{table}

Before discussing the EPR paradox, let us introduce some terminology. When
focusing on a realistic local hidden-variable theory, we emphasize that: i)
\emph{realism} means that physical systems have objective properties that are
defined prior to, and independent of, observation; ii) \emph{locality} means
Einstein locality, that is, physical influences do not propagate faster than
light (i.e., superluminal communication is not possible); iii)
\emph{hidden-variables} are variables that are not accessible from an
empirical standpoint. For completeness, we also remark that \emph{quantum
nonlocality} denotes the possibility that, thanks to entanglement,
correlations among properties of a physical system can propagate
instantaneously. Finally, \emph{Bell locality} is a property that means the
factorization of the joint probability distribution of the properties of the
two subsystems that specify a composite physical system. The coexistence of
realism and locality within the same theoretical construct is at the roots of
the EPR paradox. In Table I, we report a schematic description of the meaning
of key concepts that appear in the Bell inequality discussion: Realism,
Einstein locality, Bell locality and, finally, quantum nonlocality.

In what follows, we discuss the EPR paradox \cite{epr} by following to a great
extent the pedagogical presentation in Ref. \cite{peres95}. In 1935, Einstein,
Podolsky, and Rosen focused their attention on the physics of a composite
quantum-mechanical system specified by two distant particles and described by
an entangled wave function $\psi$ defined as \cite{epr}
\begin{equation}
\psi=\psi\left(  x_{1}\text{, }x_{2}\text{; }p_{1}\text{, }p_{2}\right)
\overset{\text{def}}{=}\delta\left(  x_{1}-x_{2}-L\right)  \delta\left(
p_{1}+p_{2}\right)  . \label{wave1}%
\end{equation}
The quantity $\delta$ in Eq. (\ref{wave1}) is a normalizable function
characterized by an arbitrarily high and narrow peak. The quantity $L$,
instead, denotes a distance that is much larger than the range of mutual
interaction between the two particles. The wave function $\psi$ in Eq.
(\ref{wave1}) has a clear physical interpretation. It specifies two particles
that have been prepared in such a way that their total momentum is arbitrarily
close to zero, and their relative distance is arbitrarily close to $L$.
Observe that the operators $x_{1}-x_{2}$ and $p_{1}+p_{2}$ in Eq.
(\ref{wave1}) commute, $\left[  x_{1}-x_{2}\text{, }p_{1}+p_{2}\right]  =0$,
since the canonical quantum-mechanical commutation relation is given by
$\left[  x_{k}\text{, }p_{l}\right]  =i\hbar\delta_{kl}$. In the state $\psi$
in Eq. (\ref{wave1}) , one does not know anything about the positions of the
individual particles. Instead, one only knows their distance from each other.
Furthermore, one does not know anything of their individual momenta and only
has proper knowledge of their total momentum. However, one shall be capable of
predicting with certainty the value of $x_{2}$ if one measures $x_{1}$,
without having disturbed particle $2$ in any way. Therefore, EPR argue that no
real change can happen in the second system in consequence of anything that
may be performed on the first system since at the time of measurement the two
systems do not interact any longer. Therefore, according to EPR, $x_{2}$
corresponds to an element of physical reality. Alternatively, if one desires
to perform a measurement of $p_{1}$ rather than $x_{1}$, one shall then be
capable of predicting with certainty the value of $p_{2}$ without having
perturbed particle $2$ in any way. Therefore, by the same argument as above,
$p_{2}$ also is an element of reality according to EPR. From this reasoning,
it appears that $x_{2}$ and $p_{2}$ are equally elements of reality. However,
\ since $\left[  x_{2}\text{, }p_{2}\right]  \neq0$, these operators do not
commute and quantum mechanics forbids the simultaneous assignment of precise
values to both $x_{2}$ and $p_{2}$. For this reason, EPR are forced to arrive
at the conclusion that the quantum-mechanical description of physical reality
in terms of wave functions is incomplete. At the same time, EPR cautiously
leave open the possibility of the existence of a complete theory. According to
EPR, reality can be described as follows: If one can predict with certainty
the value of a physical quantity without perturbing the system in any way,
then there is an element of reality that corresponds to such a physical
quantity. The EPR article was not incorrect and, arguably, it had been written
too early as pointed out by Peres in Ref. \cite{peres04}. The EPR argument did
not consider that, like any other physical object, the observer's information
was localized. Specifically, in addition to being an abstract notion,
information needs as approximately localized physical carrier. The problem is
that Einstein's locality principle states that events happening in a given
spacetime region do not depend on external parameters that may be tuned, at
the same moment, by agents placed in distant spacetime regions. However, in
quantum physics, one has to accept that a measurement on a part of the system
is to be considered as a measurement on the whole system. If one wishes to
keep Einstein's locality principle, alternative theories that incorporate such
a principle lead to a testable inequality constraint among certain observables
that violates quantum-mechanical predictions. This is the so-called Bell's
inequality \cite{bell64}. From an experimental standpoint, several violations
of Bell's inequality have been observed \cite{peres95}. Therefore, despite
that fact that this situation may appear psychologically uncomfortable,
quantum mechanics has prevailed over alternative theories because of the
experimental verdict. Quantum-mechanical predictions violate Bell's
inequality. Moreover, alternative theories fulfilling Einstein's locality
principle yield experimentally verifiable differences with respect to quantum
mechanics. To a certain extent, it is ironic that Bell's theorem can be
regarded as one of the \ most profound discovery of science because it is not
obeyed by experimental facts \cite{stapp75}. We emphasize that Bell's work in
Ref. \cite{bell64} is not about quantum mechanics. Rather, it is a general
proof where it is shown the existence of an upper limit to the correlation of
distant events in any physical theory that assumes the validity of Einstein's
principle of local causes. In particular, Bell showed that in a theory where
parameters are introduced to obtain the results of individual measurements,
without modifying the statistical predictions, there must exist a mechanism
whereby the setting of one measuring apparatus can affect the reading of
another instrument, however remote. Furthermore, such a theory could not be
Lorenz invariant since the signal being considered must propagate in an
instantaneous fashion. In particular, given an arbitrary entangled quantum
state $\left\vert \psi\right\rangle $, one can find pairs of observables whose
correlations violate Bell's inequality (i.e., Eq. (\ref{bell})). This implies
that, for such a state $\left\vert \psi\right\rangle $, quantum mechanics
makes statistical predictions that are not compatible with Einstein's locality
(i.e., the principle of local causes). Locality requires that the outcomes of
experiments carried out at a given location in space are not dependent on
arbitrary choices of other experiments that can be performed, simultaneously,
at distant locations in space. Bell's theorem leads to the conclusion that
quantum mechanics is not compatible with the viewpoint that physical
observables have pre-existing values that do not depend on the measurement
process. A hidden variable theory that incorporates Einstein's locality
principle would predict individual events violating the canonical principle of
special relativity. Namely, there would exist no covariant discrimination
between cause and effect. It turns out that the EPR paradox is settled in a
manner which Einstein would have disliked the most. This point shall be
further addressed in the next sections. For the time being, we report two
different realizations of the EPR gedankenexperiment in Table II. Having
introduced the EPR paradox, we are ready to discuss the derivation of \ Bell's
inequality in a RLHVT.\begin{table}[t]
\centering
\begin{tabular}
[c]{c|c|c|c}\hline\hline
\textbf{Scientists} & \textbf{Particle pairs} & \textbf{Experimental scheme} &
\textbf{Quantity of interest}\\\hline
Bell, 1962 & Electrons & Stern-Gerlach magnet & Spin correlation coefficient\\
Aspect-Grangier-Roger, 1982 & Photons & Two-channel polarizer & Polarization
correlation coefficient\\\hline
\end{tabular}
\caption{Different realizations of the Einstein-Podolski-Rosen (EPR)
gedankenexperiment. The spin and polarization correlation coefficients can be
experimentally determined in terms of coincidence counting rates of spin and
polarization configurations, respectively.}%
\end{table}

\section{Derivation of Bell's inequality in a RLHVT}

In this section, we present the derivation of the Bell inequality in the 1969
Clause-Horne-Shimony-Holt (CHSH) form (see Eq. (1a) in Ref. \cite{clauser69})
as originally discussed in 1971 by Bell (see Eqs. (9) and (20) in Chapters 4
and 16, respectively, of Ref. \cite{bell04}). This inequality was derived
within a theoretical framework of a realistic local hidden variable theory
(RLHVT). Before entering the discussion of the derivation, let us explain in a
clear manner the significance of \emph{Bell's locality}. Let us consider two
distant systems $S_{A}$ and $S_{B}$. Let $\rho_{AB}\left(  x_{A}\text{, }%
x_{B}\left\vert X_{A}\text{, }X_{B}\text{, }\lambda\right.  \right)  $ be the
joint probability distribution of the measurement outcomes $x_{A}$ and $x_{B}$
of the properties $X_{A}$ and $X_{B}$ corresponding to systems $A$ and $B$,
respectively%
\begin{equation}
\rho_{AB}\left(  x_{A}\text{, }x_{B}\left\vert X_{A}\text{, }X_{B}\text{,
}\lambda\right.  \right)  =\rho_{A}\left(  x_{A}\left\vert x_{B}\text{, }%
X_{A}\text{, }X_{B}\text{, }\lambda\right.  \right)  \rho_{B}\left(
x_{B}\left\vert X_{A}\text{, }X_{B}\text{, }\lambda\right.  \right)  \text{.}
\label{jointE}%
\end{equation}
The quantity $\lambda$ specifies the classical hidden-variable model. Then,
the Einstein locality condition implies that the measurement outcomes $x_{A}$
($x_{B}$) at system $S_{A}$ ($S_{B}$) cannot be influenced by the selected
property $X_{B}$ ($X_{A}$) being measured on the second system $S_{B}$
($S_{A}$). Therefore, the Einstein locality implies that $\rho_{A}\left(
x_{A}\left\vert x_{B}\text{, }X_{A}\text{, }X_{B}\text{, }\lambda\right.
\right)  =\rho_{A}\left(  x_{A}\left\vert X_{A}\text{, }\lambda\right.
\right)  $ and $\rho_{B}\left(  x_{B}\left\vert X_{A}\text{, }X_{B}\text{,
}\lambda\right.  \right)  =\rho_{B}\left(  x_{B}\left\vert X_{B}\text{,
}\lambda\right.  \right)  $. Therefore, the Einstein locality implies that the
joint probability distribution in Eq. (\ref{jointE}) can be recast in a
separable form as%
\begin{equation}
\rho_{AB}\left(  x_{A}\text{, }x_{B}\left\vert X_{A}\text{, }X_{B}\text{,
}\lambda\right.  \right)  =\rho_{A}\left(  x_{A}\left\vert X_{A}\text{,
}\lambda\right.  \right)  \rho_{B}\left(  x_{B}\left\vert X_{B}\text{,
}\lambda\right.  \right)  \text{.} \label{jointB}%
\end{equation}
Eq. (\ref{jointB}) is known as the Bell locality condition. As a final remark,
we observe that the Bell locality condition is a weaker form of the Einstein
locality condition since the latter implies the former (and not the converse).
We are now ready to discuss the derivation.

Consider a system of two spin-$1/2$ particles prepared in a state such that
they proceed in distinct directions towards two measuring apparatuses.\ Assume
these apparatuses are used to measure spin components along unit directions
$\hat{a}$ and $\hat{b}$. Furthermore, assume the hypothetical complete
direction of the initial state of the composite quantum system can be
expressed by means of a hidden variable $\lambda$ distributed according to a
probability distribution $\rho\left(  \lambda\right)  $ satisfying the
normalization condition $\int\rho\left(  \lambda\right)  d\lambda=1$. From our
previous discussion, we note that locality requires that observables $A$ and
$B$ of the composite quantum system are such that $A=A\left(  \hat{a}\text{,
}\lambda\right)  $ and $B=B\left(  \hat{b}\text{, }\lambda\right)  $. However,
$A$ ($B$) does not depend on $\hat{b}$ ($\hat{a}$) and, in addition, the
measurable results of $A$ ($B$) are supposed to be $\pm1$. Assume the spin
correlation coefficient $E\left(  \hat{a}\text{, }\hat{b}\right)  $ in this
local realistic theory is given by,%
\begin{equation}
E\left(  \hat{a}\text{, }\hat{b}\right)  \overset{\text{def}}{=}\int A\left(
\hat{a}\text{, }\lambda\right)  B\left(  \hat{b}\text{, }\lambda\right)
\rho\left(  \lambda\right)  d\lambda\text{.} \label{SCC1}%
\end{equation}
More generally, the measuring devices themselves could depend on hidden
variables which could affect the experimental results. Averaging first over
these variables, $E\left(  \hat{a}\text{, }\hat{b}\right)  $ in Eq.
(\ref{SCC1}) can be formally represented as \cite{bell04}
\begin{equation}
E\left(  \hat{a}\text{, }\hat{b}\right)  \overset{\text{def}}{=}\int\bar
{A}\left(  \hat{a}\text{, }\lambda\right)  \bar{B}\left(  \hat{b}\text{,
}\lambda\right)  \rho\left(  \lambda\right)  d\lambda\text{,} \label{scc2}%
\end{equation}
with $\left\vert \bar{A}\right\vert \leq1$ and $\left\vert \bar{B}\right\vert
\leq1$. Let $\hat{a}^{\prime}$ and $\hat{b}^{\prime}$ be alternative
orientations of the apparatuses. Then, we have
\begin{equation}
E\left(  \hat{a}\text{, }\hat{b}\right)  -E\left(  \hat{a}\text{, }\hat
{b}^{\prime}\right)  =\int\left[  \bar{A}\left(  \hat{a}\text{, }%
\lambda\right)  \bar{B}\left(  \hat{b}\text{, }\lambda\right)  -\bar{A}\left(
\hat{a}\text{, }\lambda\right)  \bar{B}\left(  \hat{b}^{\prime}\text{,
}\lambda\right)  \right]  \rho\left(  \lambda\right)  d\lambda\text{,}%
\end{equation}
that is, adding and subtracting suitably chosen identical terms,%
\begin{equation}
E\left(  \hat{a}\text{, }\hat{b}\right)  -E\left(  \hat{a}\text{, }\hat
{b}^{\prime}\right)  =\int\left\{
\begin{array}
[c]{c}%
\bar{A}\left(  \hat{a}\text{, }\lambda\right)  \bar{B}\left(  \hat{b}\text{,
}\lambda\right)  \left[  1\pm\bar{A}\left(  \hat{a}^{\prime}\text{, }%
\lambda\right)  \bar{B}\left(  \hat{b}^{\prime}\text{, }\lambda\right)
\right]  +\\
-\bar{A}\left(  \hat{a}\text{, }\lambda\right)  \bar{B}\left(  \hat{b}%
^{\prime}\text{, }\lambda\right)  \left[  1\pm\bar{A}\left(  \hat{a}^{\prime
}\text{, }\lambda\right)  \bar{B}\left(  \hat{b}\text{, }\lambda\right)
\right]
\end{array}
\right\}  \rho\left(  \lambda\right)  d\lambda\text{.} \label{scc3}%
\end{equation}
Recalling that $\left\vert x-y\right\vert \leq\left\vert x\right\vert
+\left\vert y\right\vert $ for any $x$, $y\in%
\mathbb{R}
$ and, in addition, $\left\vert x\right\vert =x$ when $x\geq0$, Eq.
(\ref{scc3}) yields%
\begin{align}
\left\vert E\left(  \hat{a}\text{, }\hat{b}\right)  -E\left(  \hat{a}\text{,
}\hat{b}^{\prime}\right)  \right\vert  &  =\left\vert \int\left\{
\begin{array}
[c]{c}%
\bar{A}\left(  \hat{a}\text{, }\lambda\right)  \bar{B}\left(  \hat{b}\text{,
}\lambda\right)  \left[  1\pm\bar{A}\left(  \hat{a}^{\prime}\text{, }%
\lambda\right)  \bar{B}\left(  \hat{b}^{\prime}\text{, }\lambda\right)
\right]  +\\
-\bar{A}\left(  \hat{a}\text{, }\lambda\right)  \bar{B}\left(  \hat{b}%
^{\prime}\text{, }\lambda\right)  \left[  1\pm\bar{A}\left(  \hat{a}^{\prime
}\text{, }\lambda\right)  \bar{B}\left(  \hat{b}\text{, }\lambda\right)
\right]
\end{array}
\right\}  \rho\left(  \lambda\right)  d\lambda\right\vert \nonumber\\
&  \leq\left\vert \int\bar{A}\left(  \hat{a}\text{, }\lambda\right)  \bar
{B}\left(  \hat{b}\text{, }\lambda\right)  \left[  1\pm\bar{A}\left(  \hat
{a}^{\prime}\text{, }\lambda\right)  \bar{B}\left(  \hat{b}^{\prime}\text{,
}\lambda\right)  \right]  \rho\left(  \lambda\right)  d\lambda\right\vert
+\nonumber\\
&  +\left\vert \int\bar{A}\left(  \hat{a}\text{, }\lambda\right)  \bar
{B}\left(  \hat{b}^{\prime}\text{, }\lambda\right)  \left[  1\pm\bar{A}\left(
\hat{a}^{\prime}\text{, }\lambda\right)  \bar{B}\left(  \hat{b}\text{,
}\lambda\right)  \right]  \rho\left(  \lambda\right)  d\lambda\right\vert
\nonumber\\
&  \leq\left\vert \int\left[  1\pm\bar{A}\left(  \hat{a}^{\prime}\text{,
}\lambda\right)  \bar{B}\left(  \hat{b}^{\prime}\text{, }\lambda\right)
\right]  \rho\left(  \lambda\right)  d\lambda\right\vert +\left\vert
\int\left[  1\pm\bar{A}\left(  \hat{a}^{\prime}\text{, }\lambda\right)
\bar{B}\left(  \hat{b}\text{, }\lambda\right)  \right]  \rho\left(
\lambda\right)  d\lambda\right\vert \nonumber\\
&  =2\pm\left[  E\left(  \hat{a}^{\prime}\text{, }\hat{b}^{\prime}\right)
+E\left(  \hat{a}^{\prime}\text{, }\hat{b}\right)  \right]  \text{,}%
\end{align}
that is,%
\begin{equation}
\left\vert E\left(  \hat{a}\text{, }\hat{b}\right)  -E\left(  \hat{a}\text{,
}\hat{b}^{\prime}\right)  \right\vert \leq2\pm\left[  E\left(  \hat{a}%
^{\prime}\text{, }\hat{b}^{\prime}\right)  +E\left(  \hat{a}^{\prime}\text{,
}\hat{b}\right)  \right]  \text{.} \label{scc4}%
\end{equation}
Finally, following Ref. \cite{bell04}, an alternative and more symmetric way
to rewrite the inequality in Eq. (\ref{scc4}) is
\begin{equation}
\left\vert E\left(  \hat{a}\text{, }\hat{b}\right)  -E\left(  \hat{a}\text{,
}\hat{b}^{\prime}\right)  \right\vert +\left\vert E\left(  \hat{a}^{\prime
}\text{, }\hat{b}^{\prime}\right)  +E\left(  \hat{a}^{\prime}\text{, }\hat
{b}\right)  \right\vert \leq2\text{.} \label{CHSH}%
\end{equation}
The inequality in Eq. (\ref{CHSH}), also reported in Table III, is the
well-known CHSH inequality that appears in Ref. \cite{clauser69}. Observe that
identifying the hidden variable theory quantity $E\left(  \hat{a}\text{, }%
\hat{b}\right)  $ in Eq. (\ref{scc2}) with the quantum-mechanical quantity
$P\left(  a\text{, }b\right)  $ in Eq. (\ref{quantity}), violation of Eq.
(\ref{CHSH}) yields Eq. (\ref{bell}) (i.e., the violation of Bell's inequality
in QM). As a final remark, note that when $\hat{a}^{\prime}=\hat{b}^{\prime}$
in Eq. (\ref{CHSH}), assuming $E\left(  \hat{b}^{\prime}\text{, }\hat
{b}^{\prime}\right)  =-1$, Eq. (\ref{CHSH}) reduces to%
\begin{equation}
\left\vert E\left(  \hat{a}\text{, }\hat{b}\right)  -E\left(  \hat{a}\text{,
}\hat{b}^{\prime}\right)  \right\vert \leq2-\left\vert -1+E\left(  \hat
{b}^{\prime}\text{, }\hat{b}\right)  \right\vert \text{,}%
\end{equation}
that is,%
\begin{equation}
\left\vert E\left(  \hat{a}\text{, }\hat{b}\right)  -E\left(  \hat{a}\text{,
}\hat{b}^{\prime}\right)  \right\vert \leq1+E\left(  \hat{b}^{\prime}\text{,
}\hat{b}\right)  \text{.} \label{bello}%
\end{equation}
The inequality in Eq. (\ref{bello}) is the original result obtained by Bell in
1964 (see Eq. (15) in Ref. \cite{bell64}). From Eq. (\ref{CHSH}), we note that
the CHSH Bell inequality is obtained within the framework of a RLHVT, applies
to a pair of two-state widely separated systems and, finally, \ constrains the
statistics of the measurement outcomes in terms of the value of a linear
combination of four correlation functions between the two systems. We are now
ready to show from a theoretical standpoint that QM violates the CHSH Bell
inequality in Eq. (\ref{CHSH}).\begin{table}[t]
\centering
\begin{tabular}
[c]{c}\hline\hline
\textbf{The Bell inequality in the CHSH form}\\\hline
$\left\vert E\left(  \hat{a}\text{, }\hat{b}\right)  -E\left(  \hat{a}\text{,
}\hat{b}^{\prime}\right)  \right\vert +\left\vert E\left(  \hat{a}^{\prime
}\text{, }\hat{b}^{\prime}\right)  +E\left(  \hat{a}^{\prime}\text{, }\hat
{b}\right)  \right\vert \leq2$\\\hline
\end{tabular}
\caption{The Bell inequality in the Clause-Horne-Shimony-Holt (CHSH) form.
Experimentally, the inequality demands four measurements, one for each and
every polarizaton correlation coefficient: $E\left(  \hat{a}\text{, }\hat
{b}\right)  $, $E\left(  \hat{a}\text{, }\hat{b}^{\prime}\right)  $, $E\left(
\hat{a}^{\prime}\text{, }\hat{b}^{\prime}\right)  $, and $E\left(  \hat
{a}^{\prime}\text{, }\hat{b}\right)  $. In particular, $E\left(  \hat
{a}\text{, }\hat{b}\right)  $ requires an experimental setup with two
polarizers in orientations $\hat{a}$ and $\hat{b}$.}%
\end{table}

\section{Violation of Bell's inequality in QM}

This section is partitioned in two subsections. In the first subsection, we
explicitly show that QM violates the CHSH Bell inequality when the physical
systems are in entangled quantum states. In the second subsection, we
illustrate with a simple set of examples how the degree of this violation
depends on the orientation of the polarizers and, in addition, on the degree
of entanglement of the quantum states specifying the physical systems being investigated.

\subsection{Formal discussion}

In this first subsection, we revisit Bell's theorem as presented by Gisin in
Ref. \cite{gisin91}. Moreover, we refer to Ref. \cite{gisin92} for a more
general derivation of Bell's theorem by Gisin and Peres. Gisin's version of
Bell's Theorem can be described as follows \cite{gisin91}:\ Let $\left\vert
\psi\right\rangle $ be a two-particles quantum state that belongs to the
composite Hilbert space $\mathcal{H}_{1}\otimes\mathcal{H}_{2}$. Then,
$\left\vert \psi\right\rangle $ violates Bell's inequality if $\left\vert
\psi\right\rangle $ is entangled (i.e., nonfactorable). Stated otherwise and
using standard quantum-mechanical notations, there exist quantum-mechanical
projectors $a$, $a^{\prime}$, $b$, and $b^{\prime}$ with $a\overset{\text{def}%
}{=}\left(  1/2\right)  \left[  \mathrm{I}+\hat{a}\cdot\vec{\sigma}\right]  $
such that%
\begin{equation}
\left\vert P\left(  a\text{, }b\right)  -P\left(  a\text{, }b^{\prime}\right)
\right\vert +P\left(  a^{\prime}\text{, }b\right)  +P\left(  a^{\prime}\text{,
}b^{\prime}\right)  >2\text{.} \label{bell}%
\end{equation}
For completeness, we remark that $\mathrm{I}$ is the identity operator on the
single-qubit Hilbert space, $\hat{a}$ is a unit vector, and $\vec{\sigma}$ is
the three-dimensional vector of Pauli operators. Eq. (\ref{bell}) is Bell's
inequality in the so-called Clauser-Horne-Shimony-Holt (CHSH) form
\cite{clauser69}. The quantity $P\left(  a\text{, }b\right)  $ in Eq.
(\ref{bell}) is the quantum-mechanical analog of $E\left(  \hat{a}\text{,
}\hat{b}\right)  $ in Eq. (\ref{scc2}) and represents the spin correlation
coefficient given by%
\begin{equation}
P\left(  a\text{, }b\right)  \overset{\text{def}}{=}\left\langle \left(
2a-1\right)  \otimes\left(  2b-1\right)  \right\rangle _{\psi}\text{.}
\label{quantity}%
\end{equation}
To verify the satisfaction of the inequality in Eq. (\ref{bell}), we introduce
first some relevant preliminary remarks.

\begin{enumerate}
\item[{[i]}] First, the Schmidt decomposition theorem \cite{mosca} yields the
following result. If $\left\vert \psi\right\rangle \in$ $\mathcal{H}%
_{1}\otimes\mathcal{H}_{2}$, then there is an orthonormal basis $\left\{
\left\vert \varphi_{l}\right\rangle \right\}  _{1\leq l\leq n_{\mathcal{H}%
_{1}}}$ of $\mathcal{H}_{1}$ with $n_{\mathcal{H}_{1}}\overset{\text{def}%
}{=}\dim_{%
\mathbb{C}
}\mathcal{H}_{1}$, and an orthonormal basis $\left\{  \left\vert \phi
_{m}\right\rangle \right\}  _{1\leq m\leq n_{\mathcal{H}_{2}}}$ of
$\mathcal{H}_{2}$ with $n_{\mathcal{H}_{2}}\overset{\text{def}}{=}\dim_{%
\mathbb{C}
}\mathcal{H}_{2}$, and non-negative \emph{real} numbers $\left\{
c_{k}\right\}  $ so that%
\begin{equation}
\left\vert \psi\right\rangle =\sum_{k}c_{k}\left\vert \varphi_{k}\right\rangle
\otimes\left\vert \phi_{k}\right\rangle \text{.} \label{state}%
\end{equation}
The quantities $c_{k}$ in Eq. (\ref{state}) are termed Schmidt coefficients,
and the number of terms in the summation is at most the minimum of
$n_{\mathcal{H}_{1}}$ and $n_{\mathcal{H}_{2}}$. To consider an entangled
state of two-particle systems, one can assume with no loss of generality that
at least two $c_{k}$-coefficients are nonvanishing in Eq. (\ref{state}).
Assume, for instance, $c_{1}\neq0\neq c_{2}$ and $c_{k}=0$ for any $k>2$. For
completeness, we point out that in principle $c_{k}$-coefficients with $k>2$
could also be different from zero. However, with no loss of generality, one
can focus the discussion to the case in which $c_{k}=0$ for any $k>2$, since
it can be shown that the Bell inequality in the CHSH form \cite{clauser69} can
be violated in any two-dimensional Hilbert subspace with nonzero Schmidt
coefficients \cite{bell64, prlgisin}. For completeness, we point out that the
projection of the joint state of n pairs of particles onto a subspace spanned
by states having a common Schmidt coefficients is a key step in the so-called
Schmidt projection method, a technique used to study entanglement
concentration in any pure state of a bipartite system \cite{fuck01}.
Furthermore, for a practical demonstration of the violation of the CHSH
version of Bell's inequality in $2D$ subspaces of an higher-dimensional
orbital angular momentum Hilbert space, we refer to \cite{fuck02}. For
completeness, we emphasize at this point that an essential step in the
so-called Schmidt projection method, a technique employed to investigate
entanglement concentration in arbitrary pure states of bipartite quantum
systems \cite{fuck01}, is the projection of the joint state of $n$-pairs of
particles onto a subspace spanned by states that share a common set of Schmidt
coefficients. Moreover, we refer to Ref. \cite{fuck02} for a verification of
the violation of the Bell inequality in the CHSH form in two-dimensional
subspaces of an higher-dimensional orbital angular momentum Hilbert space.

\item[{[ii]}] Second, the measure of entanglement vanishes for any separable
quantum state. Moreover, the degree of entanglement of a composite quantum
state does not change under local unitary transformations \cite{vedral}. A
local unitary transformation is essentially a change of basis with respect to
which we consider a given entangled state. Since, at any given time, one could
just reverse the basis change, a change of basis should not be responsible for
modifying the amount of entanglement accessible to an external observer.
Therefore, the amount of entanglement should be the same in both bases. For
convenience and with no change in the entanglement behavior of the quantum
state $\left\vert \psi\right\rangle $, given these above-mentioned comments,
let us apply a local unitary transformation $\left(  U_{1}\otimes
U_{2}\right)  $ on $\left\vert \psi\right\rangle $ such that%
\begin{equation}
\left\vert \psi\right\rangle \rightarrow\left(  U_{1}\otimes U_{2}\right)
\left\vert \psi\right\rangle =\left(  U_{1}\otimes U_{2}\right)  \left(
c_{1}\left\vert \varphi_{1}\right\rangle \otimes\left\vert \phi_{1}%
\right\rangle +c_{2}\left\vert \varphi_{2}\right\rangle \otimes\left\vert
\phi_{2}\right\rangle \right)  \text{,}%
\end{equation}
where,%
\begin{align}
\left(  U_{1}\otimes U_{2}\right)  \left(  c_{1}\left\vert \varphi
_{1}\right\rangle \otimes\left\vert \phi_{1}\right\rangle +c_{2}\left\vert
\varphi_{2}\right\rangle \otimes\left\vert \phi_{2}\right\rangle \right)   &
=c_{1}U_{1}\left\vert \varphi_{1}\right\rangle \otimes U_{2}\left\vert
\phi_{1}\right\rangle +c_{2}U_{1}\left\vert \varphi_{2}\right\rangle \otimes
U_{2}\left\vert \phi_{2}\right\rangle \nonumber\\
&  =c_{1}\left\vert 01\right\rangle +c_{2}\left\vert 10\right\rangle \text{.}
\label{sopra1}%
\end{align}
In the specific scenario of spin-$1/2$ systems, the state $\left\vert
0\right\rangle $ and $\left\vert 1\right\rangle $ in\ Eq. (\ref{sopra1}) are
defined as $\left\vert 0\right\rangle \overset{\text{def}}{=}\binom{1}{0}$,
and $\left\vert 1\right\rangle \overset{\text{def}}{=}\binom{0}{1}$, respectively.

\item[{[iii]}] Third, an arbitrary density matrix $\rho$ for a single-qubit
quantum system \ can be recast as $\rho\overset{\text{def}}{=}(1/2)\left[
\mathrm{I}+\vec{p}\cdot\vec{\sigma}\right]  $ \cite{nielsen2000, cafaro2012},
where $\vec{p}\in%
\mathbb{R}
^{3}$ with $\left\Vert \vec{p}\right\Vert \leq1$ is the Bloch vector
corresponding to the state $\rho$, $\mathrm{I}$ is the $\left(  2\times
2\right)  $-identity matrix , and $\vec{\sigma}$ is the Pauli matrix vector
given in an explicit fashion by
\cite{nielsen2000,cafaro2010,cafaro2014,cafaro2022},%
\begin{equation}
\vec{\sigma}=\left[  \sigma_{x}\text{, }\sigma_{y}\text{, }\sigma_{z}\right]
\overset{\text{def}}{\mathbf{=}}\left[  \left(
\begin{array}
[c]{cc}%
0 & 1\\
1 & 0
\end{array}
\right)  \text{, }\left(
\begin{array}
[c]{cc}%
0 & -i\\
i & 0
\end{array}
\right)  \text{, }\left(
\begin{array}
[c]{cc}%
1 & 0\\
0 & -1
\end{array}
\right)  \right]  \text{.} \label{pauli}%
\end{equation}
Note that a state $\rho$ is pure if and only if $\left\Vert \vec{p}\right\Vert
=1$. In addition, a density matrix for a pure state can be described as
$\rho=\left\vert \psi\right\rangle \left\langle \psi\right\vert $ with
$\left\vert \psi\right\rangle \left\langle \psi\right\vert $ being an
orthogonal projector since $\rho^{2}=\rho$ and $\rho^{\dagger}=\rho$. The
symbol \textquotedblleft$\dagger$\textquotedblright\ is used to denote the
usual Hermitian conjugation operation in quantum mechanics.
\end{enumerate}

Taking into consideration the remarks presented in the three points [i], [ii],
and [iii], we aim at verifying the correctness of the inequality in Eq.
(\ref{bell}) for entangled states $\left\vert \psi\right\rangle $ in Eq.
(\ref{state}) and for a convenient selection of projectors $a$, $a^{\prime}$,
$b$, and $b^{\prime}$. The projectors $a$, $a^{\prime}$, $b$, $b^{\prime}$ are
given by,%
\begin{equation}
a\overset{\text{def}}{=}\frac{1+\hat{a}\cdot\vec{\sigma}}{2}\text{, }%
a^{\prime}\overset{\text{def}}{=}\frac{1+\hat{a}^{\prime}\cdot\vec{\sigma}}%
{2}\text{, }b\overset{\text{def}}{=}\frac{1+\hat{b}\cdot\vec{\sigma}}%
{2}\text{, }b^{\prime}\overset{\text{def}}{=}\frac{1+\hat{b}^{\prime}\cdot
\vec{\sigma}}{2}\text{,} \label{projectors}%
\end{equation}
with $\hat{a}$, $\hat{a}^{\prime}$, $\hat{b}$, $\hat{b}^{\prime}\in%
\mathbb{R}
^{3}$ being unit vectors so that $a^{2}=a$, $a^{\prime2}=a^{\prime}$,
$b^{2}=b$, and $b^{\prime2}=b^{\prime}$. Using Eq. (\ref{projectors}), the
spin correlation coefficient $P\left(  a\text{, }b\right)  \overset{\text{def}%
}{=}\left\langle \left(  2a-1\right)  \otimes\left(  2b-1\right)
\right\rangle _{\psi}$ in Eq. (\ref{quantity}) becomes%
\begin{equation}
P\left(  a\text{, }b\right)  =\left\langle \psi|\left(  \hat{a}\cdot
\vec{\sigma}\right)  \otimes(\hat{b}\cdot\vec{\sigma})|\psi\right\rangle
\text{,} \label{gigi}%
\end{equation}
where $\left\vert \psi\right\rangle \overset{\text{def}}{=}c_{1}\left\vert
10\right\rangle +c_{2}\left\vert 01\right\rangle $ and $c_{1}$, $c_{2}\in%
\mathbb{R}
\backslash\left\{  0\right\}  $. To evaluate $P\left(  a\text{, }b\right)  $
in Eq. (\ref{gigi}), let us find the explicit expression of $\left(
\vec{\sigma}\cdot\hat{a}\right)  \otimes(\vec{\sigma}\cdot\hat{b})$. Making
use of Eq. (\ref{pauli}), we\textbf{ }get after some algebra%
\begin{equation}
\left(  \hat{a}\cdot\vec{\sigma}\right)  \otimes(\hat{b}\cdot\vec{\sigma
})=\left(
\begin{array}
[c]{cccc}%
a_{z}b_{z} & a_{z}\left(  b_{x}-ib_{y}\right)  & b_{z}\left(  a_{x}%
-ia_{y}\right)  & \left(  a_{x}-ia_{y}\right)  \left(  b_{x}-ib_{y}\right) \\
a_{z}\left(  b_{x}+ib_{y}\right)  & -a_{z}b_{z} & \left(  a_{x}-ia_{y}\right)
\left(  b_{x}+ib_{y}\right)  & -b_{z}\left(  a_{x}-ia_{y}\right) \\
b_{z}\left(  a_{x}+ia_{y}\right)  & \left(  a_{x}+ia_{y}\right)  \left(
b_{x}-ib_{y}\right)  & -a_{z}b_{z} & -a_{z}\left(  b_{x}-ib_{y}\right) \\
\left(  a_{x}+ia_{y}\right)  \left(  b_{x}+ib_{y}\right)  & -b_{z}\left(
a_{x}+ia_{y}\right)  & -a_{z}\left(  b_{x}+ib_{y}\right)  & a_{z}b_{z}%
\end{array}
\right)  \text{,} \label{gigi1}%
\end{equation}
where $\hat{a}\overset{\text{def}}{=}\left(  a_{x}\text{, }a_{y}\text{, }%
a_{z}\right)  $ and $\hat{b}\overset{\text{def}}{=}\left(  b_{x}\text{, }%
b_{y}\text{, }b_{z}\right)  $. Therefore, by means of Eqs. (\ref{gigi1}),
(\ref{gigi}), and the relation $\left\vert \psi\right\rangle
\overset{\text{def}}{=}c_{1}\left\vert 10\right\rangle +c_{2}\left\vert
01\right\rangle $, the explicit expression of $P\left(  a\text{, }b\right)  $
becomes%
\begin{equation}
P\left(  a\text{, }b\right)  =-\left(  c_{1}^{2}+c_{2}^{2}\right)  a_{z}%
b_{z}+c_{1}c_{2}\left(  2a_{x}b_{x}+2a_{y}b_{y}\right)  \text{.} \label{gigi3}%
\end{equation}
Observe that the normalization condition for the state vector $\left\vert
\psi\right\rangle $ requires $\left\langle \psi|\psi\right\rangle =1$, that is
$c_{1}^{2}+c_{2}^{2}=1$. Thus, the spin correlation coefficient $P\left(
a\text{, }b\right)  $ in Eq. (\ref{gigi3}) reduces to
\begin{equation}
P\left(  a\text{, }b\right)  =+2c_{1}c_{2}\left(  a_{x}b_{x}+a_{y}%
b_{y}\right)  -a_{z}b_{z}\text{.} \label{first correction}%
\end{equation}
This straightforward mathematical calculation that yields Eq.
(\ref{first correction}) leads to revise the incorrect sign that comes into
view in Ref. \cite{gisin91}. Following Gisin's work in Ref. \cite{gisin91}, we
adopt the same working assumption according to which suitable expressions for
the Bloch vectors $\hat{a}$ and $\hat{b}$ are given by,%
\begin{equation}
\hat{a}\overset{\text{def}}{=}\left(  a_{x}\text{, }a_{y}\text{, }%
a_{z}\right)  =\left(  \sin\alpha\text{, }0\text{, }\cos\alpha\right)  \text{,
and }\hat{b}\overset{\text{def}}{=}\left(  b_{x}\text{, }b_{y}\text{, }%
b_{z}\right)  =\left(  \sin\beta\text{, }0\text{, }\cos\beta\right)  \text{,}
\label{vector}%
\end{equation}
with $\alpha=0$, $\alpha^{\prime}=\pm\pi/2$ where the sign is the \emph{same}
as that of $c_{1}c_{2}$. Unit vectors $\hat{a}^{\prime}$ and $\hat{b}^{\prime
}$ are assumed to be given by $\hat{a}^{\prime}=\left(  \sin\alpha^{\prime
}\text{, }0\text{, }\cos\alpha^{\prime}\right)  $ and $\hat{b}^{\prime
}=\left(  \sin\beta^{\prime}\text{, }0\text{, }\cos\beta^{\prime}\right)  $,
respectively. Observe that due to the previous sign mistake that appears in
Gisin's analogue of our Eq. (\ref{first correction}), in Ref. \cite{gisin91}
it is stated that $\alpha^{\prime}=\pm\pi/2$ where the sign is the
\emph{opposite} of that of $c_{1}c_{2}$. Recalling Eq. (\ref{bell}), let us
focus on the quantity $\left\vert P\left(  a\text{, }b\right)  -P\left(
a\text{, }b^{\prime}\right)  \right\vert +P\left(  a^{\prime}\text{,
}b\right)  +P\left(  a^{\prime}\text{, }b^{\prime}\right)  $. Exploiting Eqs.
(\ref{quantity}), (\ref{projectors}), and (\ref{vector}), we get%

\begin{equation}
\left\vert P\left(  a\text{, }b\right)  -P\left(  a\text{, }b^{\prime}\right)
\right\vert =\left\vert -\cos\beta+\cos\beta^{\prime}\right\vert =\left\vert
-\left(  \cos\beta-\cos\beta^{\prime}\right)  \right\vert =\left\vert
\cos\beta-\cos\beta^{\prime}\right\vert \text{,} \label{yo1}%
\end{equation}
and,
\begin{equation}
P\left(  a^{\prime}\text{, }b\right)  +P\left(  a^{\prime}\text{, }b^{\prime
}\right)  =2c_{1}c_{2}\sin\alpha^{\prime}\left(  \sin\beta+\sin\beta^{\prime
}\right)  -\cos\alpha^{\prime}\left(  \cos\beta+\cos\beta^{\prime}\right)
\text{.} \label{yo1b}%
\end{equation}
Assuming $c_{1}c_{2}>0$ and $\alpha^{\prime}=+\pi/2$, Eq. (\ref{yo1b}) becomes%
\begin{equation}
P\left(  a^{\prime}\text{, }b\right)  +P\left(  a^{\prime}\text{, }b^{\prime
}\right)  =2\left\vert c_{1}c_{2}\right\vert \left(  \sin\beta+\sin
\beta^{\prime}\right)  \text{.}%
\end{equation}
Furthermore, taking $c_{1}c_{2}<0$ and $\alpha^{\prime}=-\pi/2$, we obtain%
\begin{equation}
P\left(  a^{\prime}\text{, }b\right)  +P\left(  a^{\prime}\text{, }b^{\prime
}\right)  =-2c_{1}c_{2}\left(  \sin\beta+\sin\beta^{\prime}\right)  \text{,}
\label{yo1c}%
\end{equation}
with $-2c_{1}c_{2}>0$. Therefore $2\left\vert c_{1}c_{2}\right\vert
=-2c_{1}c_{2}>0$, and $P\left(  a^{\prime}\text{, }b\right)  +P\left(
a^{\prime}\text{, }b^{\prime}\right)  $ in Eq.\ (\ref{yo1c}) becomes%
\begin{equation}
P\left(  a^{\prime}\text{, }b\right)  +P\left(  a^{\prime}\text{, }b^{\prime
}\right)  =2\left\vert c_{1}c_{2}\right\vert \left(  \sin\beta+\sin
\beta^{\prime}\right)  \text{.} \label{yo2}%
\end{equation}
Summing up, employing Eqs. (\ref{yo1}) and (\ref{yo2}), $\left\vert P\left(
a\text{, }b\right)  -P\left(  a\text{, }b^{\prime}\right)  \right\vert
+P\left(  a^{\prime}\text{, }b\right)  +P\left(  a^{\prime}\text{, }b^{\prime
}\right)  $ can be recast as
\begin{equation}
\left\vert P\left(  a\text{, }b\right)  -P\left(  a\text{, }b^{\prime}\right)
\right\vert +P\left(  a^{\prime}\text{, }b\right)  +P\left(  a^{\prime}\text{,
}b^{\prime}\right)  =\left\vert \cos\left(  \beta\right)  -\cos\left(
\beta^{\prime}\right)  \right\vert +2\left\vert c_{1}c_{2}\right\vert
\cdot\left[  \sin\left(  \beta\right)  +\sin\left(  \beta^{\prime}\right)
\right]  \text{.} \label{q0}%
\end{equation}
Let $\beta$ and $\beta^{\prime}$ be such that $\sin\beta>0$ and $\sin
\beta^{\prime}>0$ so that $2\left\vert c_{1}c_{2}\right\vert \left(  \sin
\beta+\sin\beta^{\prime}\right)  >0$. Let us also adopt the working condition
given by
\begin{equation}
-\cos\left(  \beta^{\prime}\right)  =\cos\left(  \beta\right)  =\left(
1+4\left\vert c_{1}c_{2}\right\vert ^{2}\right)  ^{-1/2}\text{,} \label{q1}%
\end{equation}
which, after some algebra, yields%
\begin{equation}
\sin\left(  \beta\right)  =\sin\left(  \beta^{\prime}\right)  =2\left\vert
c_{1}c_{2}\right\vert \left(  1+4\left\vert c_{1}c_{2}\right\vert ^{2}\right)
^{-1/2}\text{.} \label{q2}%
\end{equation}
Therefore, substituting Eqs. (\ref{q1}) and (\ref{q2}) into Eq. (\ref{q0}), we
finally get%
\begin{align}
\left\vert P\left(  a\text{, }b\right)  -P\left(  a\text{, }b^{\prime}\right)
\right\vert +P\left(  a^{\prime}\text{, }b\right)  +P\left(  a^{\prime}\text{,
}b^{\prime}\right)   &  =2\left(  1+4\left\vert c_{1}c_{2}\right\vert
^{2}\right)  ^{-1/2}+8\left\vert c_{1}c_{2}\right\vert ^{2}\left(
1+4\left\vert c_{1}c_{2}\right\vert ^{2}\right)  ^{-1/2}\nonumber\\
& \nonumber\\
&  =2\left(  1+4\left\vert c_{1}c_{2}\right\vert ^{2}\right)  ^{-1/2}\left(
1+4\left\vert c_{1}c_{2}\right\vert ^{2}\right) \nonumber\\
& \nonumber\\
&  =2\left(  1+4\left\vert c_{1}c_{2}\right\vert ^{2}\right)  ^{1/2}%
\nonumber\\
& \nonumber\\
&  >2\text{,}%
\end{align}
that is, $\left\vert P\left(  a\text{, }b\right)  -P\left(  a\text{,
}b^{\prime}\right)  \right\vert +P\left(  a^{\prime}\text{, }b\right)
+P\left(  a^{\prime}\text{, }b^{\prime}\right)  >2$. The derivation of this
inequality concludes our formal verification concerning the fact that QM
violates the CHSH Bell inequality.

\subsection{Illustrative examples}

In this second subsection, we discuss three distinct scenarios leading to the
violation of Bell's inequality expressed as
\begin{equation}
\left\vert P\left(  a\text{, }b\right)  -P\left(  a\text{, }b^{\prime}\right)
\right\vert +P\left(  a^{\prime}\text{, }b\right)  +P\left(  a^{\prime}\text{,
}b^{\prime}\right)  >2\text{,} \label{amore}%
\end{equation}
where the spin correlation coefficient $P\left(  a\text{, }b\right)  $ is
given by $\left\langle \psi|\left(  \hat{a}\cdot\vec{\sigma}\right)
\otimes(\hat{b}\cdot\vec{\sigma})|\psi\right\rangle $.

We begin by observing that the quantities that can be manipulated in order to
obtain the above-mentioned inequality are essentially two, the set of
polarization vectors $\left\{  \hat{a}\text{, }\hat{b}\text{, }\hat{a}%
^{\prime}\text{, }\hat{b}^{\prime}\right\}  $ and the entangled quantum state
$\left\vert \psi\right\rangle $. In our discussion, we focus our attention on
$\left\vert \psi\right\rangle =c_{1}\left\vert 01\right\rangle +c_{2}%
\left\vert 10\right\rangle $ with $c_{1}$, $c_{2}\in%
\mathbb{R}
$ such that $c_{1}^{2}+c_{2}^{2}=1$ and $2c_{1}c_{2}=\pm1$. Furthermore, we
propose three different sets of polarization vectors $\left\{  \hat{a}\text{,
}\hat{b}\text{, }\hat{a}^{\prime}\text{, }\hat{b}^{\prime}\right\}  $. Note
that, in general, the polarization vectors $\hat{a}$, $\hat{b}$, $\hat
{a}^{\prime}$, $\hat{b}^{\prime}$ can be parametrized in terms of the
spherical angles $\theta\in\left[  0\text{, }\pi\right]  $ and $\varphi
\in\left[  0\text{, }2\pi\right)  $ as
\begin{align}
&  \hat{a}\overset{\text{def}}{=}\left(  \sin\left(  \theta_{1}\right)
\cos\left(  \varphi_{1}\right)  \text{, }\sin\left(  \theta_{1}\right)
\sin\left(  \varphi_{1}\right)  \text{, }\cos\left(  \theta_{1}\right)
\right)  \text{, }\hat{b}\overset{\text{def}}{=}\left(  \sin\left(  \theta
_{2}\right)  \cos\left(  \varphi_{2}\right)  \text{, }\sin\left(  \theta
_{2}\right)  \sin\left(  \varphi_{2}\right)  \text{, }\cos\left(  \theta
_{2}\right)  \right)  \text{,}\nonumber\\
&  \hat{a}^{\prime}\overset{\text{def}}{=}\left(  \sin\left(  \theta
_{3}\right)  \cos\left(  \varphi_{3}\right)  \text{, }\sin\left(  \theta
_{3}\right)  \sin\left(  \varphi_{3}\right)  \text{, }\cos\left(  \theta
_{3}\right)  \right)  \text{, and }\text{ }\hat{b}^{\prime}\overset{\text{def}%
}{=}\left(  \sin\left(  \theta_{4}\right)  \cos\left(  \varphi_{4}\right)
\text{, }\sin\left(  \theta_{4}\right)  \sin\left(  \varphi_{4}\right)
\text{, }\cos\left(  \theta_{4}\right)  \right)  \text{,} \label{B1}%
\end{align}
respectively. Clearly, considering Eq. (\ref{B1}) and Bell's inequality, we
see that in the most general case one needs to handle an inequality with eight
angular parameters, once the entangles quantum state has been chosen. To gain
physical insights with simpler configurations, in what follows we shall deal
only with two-dimensional parametric regions (i.e., regions parametrized by a
polar angle $\theta$ and an azimuthal angle $\varphi$) where the Bell
inequality is violated. This way, we shall be able to visualize this regions
with neat two-dimensional parametric plots. In particular, we shall be able to
comment on the behavior of these regions when we modify the degree of
entanglement of the quantum state $\left\vert \psi\right\rangle $ while
keeping fixed the set of polarization vectors $\left\{  \hat{a}\text{, }%
\hat{b}\text{, }\hat{a}^{\prime}\text{, }\hat{b}^{\prime}\right\}  $.

Before presenting our scenarios, we remark for completeness that the
projectors $\left\{  a\text{, }b\text{, }a^{\prime}\text{, }b^{\prime
}\right\}  $ in Eq. (\ref{amore}) need not be mutually orthogonal. Two
projectors $a\overset{\text{def}}{=}\left(  1/2\right)  \left[  \mathrm{I}%
+\hat{a}\cdot\vec{\sigma}\right]  $ and $b\overset{\text{def}}{=}\left(
1/2\right)  \left[  \mathrm{I}+\hat{b}\cdot\vec{\sigma}\right]  $ with
$a^{2}=a$ and $b^{2}=b$ are orthogonal when $ab$ is the null operator
$\mathcal{O}$. Recalling that $\left(  \hat{a}\cdot\vec{\sigma}\right)
(\hat{b}\cdot\vec{\sigma})=(\hat{a}\cdot\hat{b})\mathrm{I}+i(\hat{a}\times
\hat{b})\cdot\vec{\sigma}$, we get%
\begin{equation}
ab=\frac{1}{4}\mathrm{I}+\frac{1}{4}(\hat{a}+\hat{b})\cdot\vec{\sigma}%
+\frac{1}{4}(\hat{a}\cdot\hat{b})\mathrm{I}+\frac{1}{4}i(\hat{a}\times\hat
{b})\cdot\vec{\sigma}\text{.} \label{prodotto}%
\end{equation}
From Eq. (\ref{prodotto}), we note that $ab=\mathcal{O}$ when $\hat{b}%
=-\hat{a}$. Indeed, in this case we have $\hat{a}+\hat{b}=0$, $\hat{a}%
\cdot\hat{b}=-1$, and $\hat{a}\times\hat{b}=0$. Recasting the polarization
vectors $\hat{a}$ and $\hat{b}$ of orthogonal projectors in terms of spherical
angles $\theta$ and $\varphi$, we have $\hat{a}=\hat{a}\left(  \theta\text{,
}\varphi\right)  \overset{\text{def}}{=}\left(  \cos\left(  \varphi\right)
\sin\left(  \theta\right)  \text{, }\sin\left(  \varphi\right)  \sin\left(
\theta\right)  \text{, }\cos\left(  \theta\right)  \right)  $ and $\hat
{b}=\hat{b}\left(  \theta\text{, }\varphi\right)  \overset{\text{def}}{=}%
\hat{a}\left(  \pi-\theta\text{, }\varphi+\pi\right)  =\left(  -\cos\left(
\varphi\right)  \sin\left(  \theta\right)  \text{, }-\sin\left(
\varphi\right)  \sin\left(  \theta\right)  \text{, }-\cos\left(
\theta\right)  \right)  $ so that $\hat{a}\cdot\hat{b}=-1$. In the following
three scenarios, the set of projectors $\left\{  a\text{, }b\text{, }%
a^{\prime}\text{, }b^{\prime}\right\}  $ in Eq. (\ref{amore}) are not mutually
orthogonal. In particular, we shall focus on configurations specified by
$\hat{a}\cdot\hat{a}^{\prime}=0=\hat{b}\cdot\hat{b}^{\prime}$ and, in
addition, only two tunable angular parameters. \begin{figure}[t]
\centering
\includegraphics[width=1\textwidth] {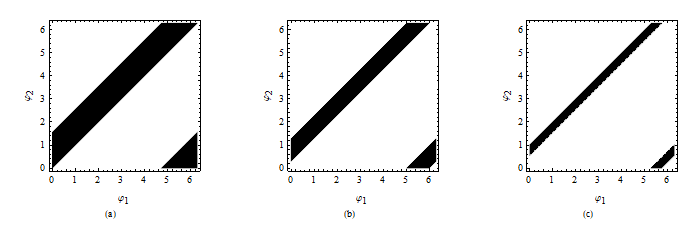}\caption{Plot of the two-dimensional
parametric region (black region) where the Bell inequality is violated. The
region is defined by the inequality $f\left(  \varphi_{1}\text{, }\varphi
_{2}\right)  >2\mathrm{C}\left(  \psi\right)  $, where $f\left(  \varphi
_{1}\text{, }\varphi_{2}\right)  \protect\overset{\text{def}}{=}\left\vert
\cos\left(  \varphi_{1}-\varphi_{2}\right)  -\sin\left(  \varphi_{1}%
-\varphi_{2}\right)  \right\vert +\left[  \cos\left(  \varphi_{1}-\varphi
_{2}\right)  -\sin\left(  \varphi_{1}-\varphi_{2}\right)  \right]  $,
$\mathrm{C}\left(  \psi\right)  \protect\overset{\text{def}}{=}2\left\vert
c_{1}c_{2}\right\vert \in\left[  0\text{, }1\right]  $, and $c_{1}c_{2}>0$. We
assume $\mathrm{C}\left(  \psi\right)  $ equals $1$, $4/5$, and $8/11$ in
plots (a), (b), and (c), respectively. Note that the parametric region tends
to shrink as the degree of entanglement $\mathrm{C}\left(  \psi\right)  $ of
the two-qubit quantum state $\left\vert \psi\right\rangle
\protect\overset{\text{def}}{=}c_{1}\left\vert 01\right\rangle +c_{2}%
\left\vert 10\right\rangle $ becomes weaker.}%
\end{figure}

\subsubsection{First scenario: $xz$-plane}

In the first scenario, we assume that polarization vectors $\left\{  \hat
{a}\text{, }\hat{b}\text{, }\hat{a}^{\prime}\text{, }\hat{b}^{\prime}\right\}
$ are in the $xz$-plane. We set $\varphi_{1}=\varphi_{2}=0$, $\varphi
_{3}=\varphi_{4}=0$, $\theta_{3}=\theta_{1}+\pi/2$, and $\theta_{4}=\theta
_{2}+\pi/2$. Then,\ Eq. (\ref{B1}) yields%
\begin{align}
&  \hat{a}\overset{\text{def}}{=}\left(  \sin\left(  \theta_{1}\right)
\text{, }0\text{, }\cos\left(  \theta_{1}\right)  \right)  \text{, }\hat
{b}\overset{\text{def}}{=}\left(  \sin\left(  \theta_{2}\right)  \text{,
}0\text{, }\cos\left(  \theta_{2}\right)  \right)  \text{,}\nonumber\\
&  \hat{a}^{\prime}\overset{\text{def}}{=}\left(  \cos\left(  \theta
_{1}\right)  \text{, }0\text{, }-\sin\left(  \theta_{1}\right)  \right)
\text{, and }\text{ }\hat{b}^{\prime}\overset{\text{def}}{=}\left(
\cos\left(  \theta_{2}\right)  \text{, }0\text{, }-\sin\left(  \theta
_{2}\right)  \right)  \text{.} \label{b1}%
\end{align}
Consider the inequality in Eq. (\ref{amore}), with $P\left(  a\text{,
}b\right)  =2c_{1}c_{2}\left(  a_{x}b_{x}+a_{y}b_{y}\right)  -a_{z}b_{z}$ and
$2c_{1}c_{2}=1$. Then, using Eq. (\ref{b1}), Eq. (\ref{amore}) becomes%
\begin{equation}
\left[  f_{\mathrm{case}\text{-}1}\left(  \theta_{1}\text{, }\theta
_{2}\right)  \right]  _{2c_{1}c_{2}=1}\overset{\text{def}}{=}\left\vert
\cos\left(  \theta_{1}+\theta_{2}\right)  +\sin\left(  \theta_{1}+\theta
_{2}\right)  \right\vert +\cos\left(  \theta_{1}+\theta_{2}\right)
+\sin\left(  \theta_{1}+\theta_{2}\right)  >2\text{.} \label{ine1}%
\end{equation}
If we set $2c_{1}c_{2}=-1$, $P\left(  a\text{, }b\right)  =-\vec{a}\cdot
\vec{b}$. Then, Eq. (\ref{amore}) becomes $\left\vert \vec{a}\cdot(\vec
{b}^{\prime}-\vec{b})\right\vert -\vec{a}^{\prime}\cdot(\vec{b}^{\prime}%
+\vec{b})>2$ and leads to the inequality%
\begin{equation}
\left[  f_{\mathrm{case}\text{-}1}\left(  \theta_{1}\text{, }\theta
_{2}\right)  \right]  _{2c_{1}c_{2}=-1}\overset{\text{def}}{=}\left\vert
\cos\left(  \theta_{1}-\theta_{2}\right)  -\sin\left(  \theta_{1}-\theta
_{2}\right)  \right\vert -\left[  \cos\left(  \theta_{1}-\theta_{2}\right)
-\sin\left(  \theta_{1}-\theta_{2}\right)  \right]  >2\text{.} \label{ine1b}%
\end{equation}
Note that both inequalities in Eq. (\ref{ine1}) and Eq. (\ref{ine1b}) could be
visualized in a two-dimensional parametric plot. Interestingly, we emphasize
that in Gisin's 1991 work as revisited in Section IV, the violation of Bell's
inequality was presented with polarization vectors $\left\{  \hat{a}\text{,
}\hat{b}\text{, }\hat{a}^{\prime}\text{, }\hat{b}^{\prime}\right\}  $ in the
$xz$-plane. Specifically, Gisin selected $\hat{a}\overset{\text{def}}{=}%
(\sin\left(  \alpha\right)  $, $0$, $\cos\left(  \alpha\right)  )$, $\hat
{b}\overset{\text{def}}{=}$ $(\sin\left(  \beta\right)  $, $0$, $\cos\left(
\beta\right)  )$, $\hat{a}^{\prime}\overset{\text{def}}{=}(\sin\left(
\alpha^{\prime}\right)  $, $0$, $\cos\left(  \alpha^{\prime}\right)  )$, and
$\hat{b}^{\prime}\overset{\text{def}}{=}$ $(\sin\left(  \beta^{\prime}\right)
$, $0$, $\cos\left(  \beta^{\prime}\right)  )$. Furthermore, it was chosen
$\alpha=0$, $\alpha^{\prime}=\pm\pi/2$ (depending on the sign of $c_{1}c_{2}%
$), and, finally, arbitrary $\beta$ and $\beta^{\prime}$.

\subsubsection{Second scenario: $xy$-plane}

In the second scenario, we assume that polarization vectors $\left\{  \hat
{a}\text{, }\hat{b}\text{, }\hat{a}^{\prime}\text{, }\hat{b}^{\prime}\right\}
$ are in the $xy$-plane. We set $\theta_{1}=\theta_{2}=\pi/2$, $\theta
_{3}=\theta_{4}=\pi/2$, $\varphi_{3}=\varphi_{1}+\pi/2$, and $\varphi
_{4}=\varphi_{2}+\pi/2$. Then,\ Eq. (\ref{B1}) yields%
\begin{align}
&  \hat{a}\overset{\text{def}}{=}\left(  \cos\left(  \varphi_{1}\right)
\text{, }\sin\left(  \varphi_{1}\right)  \text{, }0\right)  \text{, }\hat
{b}\overset{\text{def}}{=}\left(  \cos\left(  \varphi_{2}\right)  \text{,
}\sin\left(  \varphi_{2}\right)  \text{, }0\right)  \text{,}\nonumber\\
&  \hat{a}^{\prime}\overset{\text{def}}{=}\left(  -\sin\left(  \varphi
_{1}\right)  \text{, }\cos\left(  \varphi_{1}\right)  \text{, }0\right)
\text{, and }\text{ }\hat{b}^{\prime}\overset{\text{def}}{=}\left(
-\sin\left(  \varphi_{2}\right)  \text{, }\cos\left(  \varphi_{2}\right)
\text{, }0\right)  \text{.} \label{b3}%
\end{align}
Consider the inequality in (\ref{amore}). Then, using Eq. (\ref{b3}) and
setting $2c_{1}c_{2}=1$, Eq. (\ref{amore}) yields%
\begin{equation}
\left[  f_{\mathrm{case}\text{-}2}\left(  \varphi_{1}\text{, }\varphi
_{2}\right)  \right]  _{2c_{1}c_{2}=1}\overset{\text{def}}{=}\left\vert
\cos\left(  \varphi_{1}-\varphi_{2}\right)  -\sin\left(  \varphi_{1}%
-\varphi_{2}\right)  \right\vert +\cos\left(  \varphi_{1}-\varphi_{2}\right)
-\sin\left(  \varphi_{1}-\varphi_{2}\right)  >2\text{.} \label{ine2}%
\end{equation}
If we set $2c_{1}c_{2}=-1$, (\ref{ine2}) is replaced by
\begin{equation}
\left[  f_{\mathrm{case}\text{-}2}\left(  \varphi_{1}\text{, }\varphi
_{2}\right)  \right]  _{2c_{1}c_{2}=-1}\overset{\text{def}}{=}\left\vert
\cos\left(  \varphi_{1}-\varphi_{2}\right)  -\sin\left(  \varphi_{1}%
-\varphi_{2}\right)  \right\vert -\left[  \cos\left(  \varphi_{1}-\varphi
_{2}\right)  -\sin\left(  \varphi_{1}-\varphi_{2}\right)  \right]  >2\text{.}
\label{ine2b}%
\end{equation}
Again, observe that both inequalities in Eq. (\ref{ine2}) and Eq.
(\ref{ine2b}) could be visualized in a simple two-dimensional parametric plot.
Interestingly, since $P\left(  a\text{, }b\right)  =2c_{1}c_{2}\left(
a_{x}b_{x}+a_{y}b_{y}\right)  $ in this second scenario, the inequality in Eq.
(\ref{ine2}) can be extended to quantum states $\left\vert \psi\right\rangle
=c_{1}\left\vert 01\right\rangle +c_{2}\left\vert 10\right\rangle $ (where
$c_{1}$, $c_{2}\in%
\mathbb{R}
$ so that $c_{1}c_{2}>0$) with arbitrary degree of entanglement \textrm{C}%
$\left(  \psi\right)  \overset{\text{def}}{=}2\left\vert c_{1}c_{2}\right\vert
\in\left[  0\text{, }1\right]  $ as%
\begin{equation}
f\left(  \varphi_{1}\text{, }\varphi_{2}\right)  \overset{\text{def}%
}{=}\left\vert \cos\left(  \varphi_{1}-\varphi_{2}\right)  -\sin\left(
\varphi_{1}-\varphi_{2}\right)  \right\vert +\cos\left(  \varphi_{1}%
-\varphi_{2}\right)  -\sin\left(  \varphi_{1}-\varphi_{2}\right)
>2\mathrm{C}\left(  \psi\right)  \text{.} \label{meno}%
\end{equation}
A plot of the two-dimensional region where the Bell inequality expressed in
terms of Eq. (\ref{meno}) is violated appears in Fig. $1$ where we assume
\textrm{C}$\left(  \psi\right)  =1$, $4/5$, and $8/11$ in plots (a), (b), and
(c), respectively. These three \textrm{C}$\left(  \psi\right)  $-choices
correspond to three distinct selections of entangled quantum states, namely%
\begin{equation}
\left\vert \psi\right\rangle \overset{\text{def}}{=}\frac{\left\vert
01\right\rangle +\left\vert 10\right\rangle }{\sqrt{2}}\text{, }\left\vert
\psi\right\rangle \overset{\text{def}}{=}\sqrt{\frac{1}{2}+\sqrt{\frac{9}%
{100}}}\left\vert 01\right\rangle +\sqrt{\frac{1}{2}-\sqrt{\frac{9}{100}}%
}\left\vert 10\right\rangle \text{, and }\left\vert \psi\right\rangle
\overset{\text{def}}{=}\sqrt{\frac{1}{2}+\frac{\sqrt{57}}{22}}\left\vert
01\right\rangle +\sqrt{\frac{1}{2}-\frac{\sqrt{57}}{22}}\left\vert
10\right\rangle \text{,}%
\end{equation}
respectively. As expected, we observe in Fig. $1$ that as the degree of
entanglement decreases, the two-dimensional parametric region where the Bell
inequality is violated tends to become smaller. Eventually, this region
disappears completely when the two-qubit state is separable.

\subsubsection{Third scenario: $yz$-plane}

In the third scenario, we assume that polarization vectors $\left\{  \hat
{a}\text{, }\hat{b}\text{, }\hat{a}^{\prime}\text{, }\hat{b}^{\prime}\right\}
$ are in the $yz$-plane. We set $\varphi_{1}=\varphi_{2}=\pi/2$, $\varphi
_{3}=\varphi_{4}=\pi/2$, $\theta_{3}=\theta_{1}+\pi/2$, and $\theta_{4}%
=\theta_{2}+\pi/2$. Then,\ Eq. (\ref{B1}) yields%
\begin{align}
&  \hat{a}\overset{\text{def}}{=}\left(  0\text{, }\sin\left(  \theta
_{1}\right)  \text{, }\cos\left(  \theta_{1}\right)  \right)  \text{, }\hat
{b}\overset{\text{def}}{=}\left(  0\text{, }\sin\left(  \theta_{2}\right)
\text{, }\cos\left(  \theta_{2}\right)  \right)  \text{,}\nonumber\\
&  \hat{a}^{\prime}\overset{\text{def}}{=}\left(  0\text{, }\cos\left(
\theta_{1}\right)  \text{, }-\sin\left(  \theta_{1}\right)  \right)  \text{,
and }\text{ }\hat{b}^{\prime}\overset{\text{def}}{=}\left(  0\text{, }%
\cos\left(  \theta_{2}\right)  \text{, }-\sin\left(  \theta_{2}\right)
\right)  \text{.} \label{b4}%
\end{align}
Consider the inequality in (\ref{amore}). Then, using Eq. (\ref{b4}) and
putting $2c_{1}c_{2}=1$, Eq. (\ref{amore}) reduces to%
\begin{equation}
\left[  f_{\mathrm{case}\text{-}3}\left(  \theta_{1}\text{, }\theta
_{2}\right)  \right]  _{2c_{1}c_{2}=1}\overset{\text{def}}{=}\left\vert
\cos\left(  \theta_{1}+\theta_{2}\right)  +\sin\left(  \theta_{1}+\theta
_{2}\right)  \right\vert +\cos\left(  \theta_{1}+\theta_{2}\right)
+\sin\left(  \theta_{1}+\theta_{2}\right)  >2\text{.} \label{ine3}%
\end{equation}
Finally, setting $2c_{1}c_{2}=1$, (\ref{ine3}) becomes%
\begin{equation}
\left[  f_{\mathrm{case}\text{-}3}\left(  \theta_{1}\text{, }\theta
_{2}\right)  \right]  _{2c_{1}c_{2}=-1}\overset{\text{def}}{=}\left\vert
\cos\left(  \theta_{1}-\theta_{2}\right)  -\sin\left(  \theta_{1}-\theta
_{2}\right)  \right\vert -\left[  \cos\left(  \theta_{1}-\theta_{2}\right)
-\sin\left(  \theta_{1}-\theta_{2}\right)  \right]  >2\text{.} \label{ine3b}%
\end{equation}
Once again, we point out that both inequalities in Eq. (\ref{ine3}) and Eq.
(\ref{ine3b}) can be used to visualize the two-dimensional parametric region
where the Bell inequality is violated.

Having discussed the violation of Bell's inequality in the CHSH form from a
theoretical standpoint, in the next section we briefly discuss the violation
of Bell's inequality from an experimental standpoint.\begin{table}[t]
\centering
\begin{tabular}
[c]{c|c}\hline\hline
\textbf{Type of experimental loophole} & \textbf{Problem}\\\hline
Locality loophole & Excluding any communication between the polarizers\\
Detection loophole & Guaranteeing efficient measurements with no fair-sampling
assumption\\\hline
\end{tabular}
\caption{Schematic description of the two main experimental loopholes, the
locality and detection loopholes. The locality loophole requires being certain
there is no communication between the polarizers. It can be solved using
random and ultrafast switching of the orientations of the polarizers. The
detection loophole, instead, requires detecting a sufficiently large fraction
of the emitted photons in experiments with optical systems in order to avoid
the fair-sampling assumption. In experiments with electron spins, efficient
spin read-out helps avoiding the fair-sampling assumption.}%
\end{table}

\section{Violation of Bell's inequality in a laboratory}

In this section, after describing some of the most important challenges that
one finds when experimentally verifying the violation of Bell's in a
laboratory, we briefly focus on the 1982 AGR experiment where the first
experimental realization of the EPR gedankenexperiment was proposed. Finally,
we outline the links connecting RLHVTs and QM\ to experiments.

\subsection{Experimental challenges}

The path towards an experimental verification, free of loopholes, of the Bell
inequality in an EPR-type experiment has been rather interesting and
challenging from a scientific standpoint. For a review on experimental aspects
related to Bell's 1964 theorem, we refer to Ref. \cite{brunner14}. The two
main experimental loopholes in EPR-type experiments are the locality and
detection loopholes. The locality loophole demands being sure there exists no
communication between the polarizers. It can be solved employing random and
ultrafast switching of the orientations of the polarizers. The detection
loophole, instead, demands detecting a sufficiently large fraction of the
emitted photons in experiments with optical systems to avoid the fair-sampling
assumption. The solution of the detection loophole is important to obtain a
statistically significant rejection of the local-realistic null hypothesis.
Therefore, it is necessary to detect a sufficiently large fraction of the
emitted photons (i.e., avoid the fair sampling assumption) so that the
correlations of the detected photons are representative. This way, one can
completely rule out that the Bell inequality is satisfied. Solving the
locality loophole is important to make sure that the choice of setting at a
measurement site $A$ has no influence on the measurement result at site $B$.
To achieve this goal, one has to make sure that the time interval that passes
between the choice of measurement setting at site $A$ and the generation of a
measurement outcome at site $B$ is shorter than the time interval it takes for
a photon to travel between the two measurement sites. Relying on the
experience gained in several experiments and theoretical proposals
\cite{wu50,kocher67,clauser69,clauser72,aspect81}, the first experimental
realization of the EPR gedankenexperiment was performed by Aspect and
collaborators in Ref. \cite{aspect82} with a stationary switching of the
orientations of the polarizers. In Ref. \cite{aspect82B}, Aspect and
collaborators improved their experimental apparatus by means of a periodic
(yet, predictable) switching of the orientations of the polarizers. In Ref.
\cite{weihs98}, Zeilinger and collaborators observed a strong violation of
Bell's inequality in an EPR-type experiment with independent observers. They
solved the locality loophole by using a random and ultrafast switching of the
orientations of the polarizers, separated by $400$ meters \cite{aspect99}.
Finally, the first experiment where both the detection and locality loopholes
were closed simultaneously was an experiment with electron spins separated by
$1.3$ kilometres and was performed by Hensen and collaborators in Ref.
\cite{hensen15}. Finally, for a review on loopholes in Bell inequality tests
of local realism, we hint to Ref. \cite{larsson14}. In Table IV, we show a
schematic description of the two main experimental loopholes, the locality and
detection loopholes.

We are now ready to focus on the AGR experiment. \begin{figure}[t]
\centering
\includegraphics[width=0.75\textwidth] {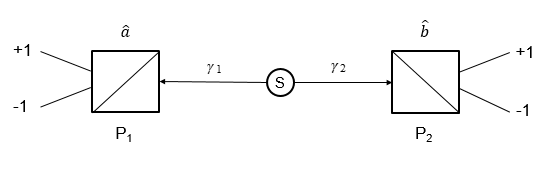}\caption{Schematic depiction of
the experimental apparatus for measuring the polarization coefficient
$E\left(  \hat{a}\text{, }\hat{b}\right)  $ in the Aspect-Grangier-Roger (AGR)
experiment in Ref. \cite{aspect82}. Two photons $\gamma_{1}$ and $\gamma_{2}$
generated by a source $\mathrm{S}$ are in a singlet state $\left\vert
\psi\right\rangle $ and propagate in opposite directions. The linear
polarizations $\left(  \pm1\right)  $ of $\gamma_{1}$ and $\gamma_{2}$ are
measured along the orientations $\hat{a}$ and $\hat{b}$ specifying a pair of
two-channel polarizers $\mathrm{P}_{\mathrm{1}}$ and $\mathrm{P}_{\mathrm{2}}%
$. The polarization coefficient $E\left(  \hat{a}\text{, }\hat{b}\right)  $ is
experimentally determined by measuring, in a single run, the four coincidence
counting rates $R_{\pm\pm}\left(  \hat{a}\text{, }\hat{b}\right)  $.}%
\end{figure}

\subsection{The 1982 AGR experiment}

In Ref. \cite{aspect82}, Aspect and collaborators proposed the first
experimental realization of the EPR gedankenexperiment. In this experiment,
they used the optical analogs of the Stern-Gerlach filters (i.e., two-channel
polarizers) to measure the linear-polarization correlation of pairs of photons
emitted in a radiative cascade of calcium. Specifically, considering that the
quantity $S_{\mathrm{RLHVT}}$ $\in\left[  -2\text{, }2\right]  $ given within
a RLHVT by%
\begin{equation}
S_{\mathrm{RLHVT}}=E(\hat{a},\hat{b})-E(\hat{a},\hat{b}^{\prime})+E(\hat
{a}^{\prime},\hat{b})+E(\hat{a}^{\prime},\hat{b}^{\prime})\text{,}%
\end{equation}
Aspect and collaborators found $S_{\mathrm{\exp}}=2.697\pm0.015$ for their
chosen set of orientations $\left\{  \hat{a}\text{, }\hat{b}\text{, }\hat
{a}^{\prime}\text{, }\hat{b}^{\prime}\right\}  $. They compared
$S_{\mathrm{\exp}}$ with $S_{\mathrm{QM}}=2.70\pm0.05$ predicted by QM. They
concluded that their experimental findings were in excellent agreement with
quantum-mechanical predictions and, thus, Bell's inequality was significantly
violated. For an in depth discussion on peculiar aspects of the experimental
data reported by Aspect while analyzing the two-channel EPR experiments in
Ref. \cite{aspect83PhD}, we refer to Refs. \cite{andrei20,andrei07}. Abnormal
features include signaling patterns in the EPR\ experimental data
\cite{andrei20} and, in addition, unexplained compensation of experimental
deviations in the coincidence counting rates leading to an experimental value
of the correlation coefficient in good agreement with quantum mechanical
predictions \cite{andrei07}. For completeness, we point out that the
correlation coefficient $E(\hat{a}$, $\hat{b})$ can be expressed, from an
experimental standpoint, as
\begin{equation}
E(\hat{a}\text{, }\hat{b})\overset{\text{def}}{=}P_{++}(\hat{a}\text{, }%
\hat{b})+P_{--}(\hat{a}\text{, }\hat{b})-P_{+-}(\hat{a}\text{, }\hat
{b})-P_{-+}(\hat{a}\text{, }\hat{b})\text{,} \label{prob1}%
\end{equation}
where the probabilities $P_{ij}(\hat{a}$, $\hat{b})$ with $i$, $j\in\left\{
+\text{, }-\right\}  $ are defined as%
\begin{equation}
P_{ij}(\hat{a}\text{, }\hat{b})\overset{\text{def}}{=}\frac{R_{ij}(\hat
{a}\text{, }\hat{b})}{R_{ii}(\hat{a}\text{, }\hat{b})+R_{jj}(\hat{a}\text{,
}\hat{b})+R_{ij}(\hat{a}\text{, }\hat{b})+R_{ji}(\hat{a}\text{, }\hat{b}%
)}\text{.} \label{prob2}%
\end{equation}
Given that a measurement along $\hat{a}$ yields the result $+1$ ($-1$) if the
polarization of the photon is found parallel (perpendicular) to $\hat{a}$,
$P_{\pm\pm}(\hat{a}$, $\hat{b})$ in Eq. (\ref{prob1}) denote the probabilities
of measuring outcomes $\pm1$ along $\hat{a}$ for particle-$1$ and $\pm1$ along
$\hat{b}$ for particle-$2$. Moreover, $R_{\pm\pm}(\hat{a}$, $\hat{b})$ in Eq.
(\ref{prob2}) denote the coincidence counting rates of measuring outcomes
$\pm1$ along $\hat{a}$ for particle-$1$ and $\pm1$ along $\hat{b}$ for
particle-$2$.

\subsection{From RLHVT and QM to experiment}

In Fig. $2$, we depict the experimental apparatus for measuring the
polarization coefficient (i.e., $E_{\mathrm{RLHVT}}(\hat{a}$, $\hat{b})$ in
Eq. (\ref{scc2})) in the Aspect-Grangier-Roger (AGR) experiment. We are now in
a good position to outline the logical steps connecting theoretical
predictions that emerge from a realistic\textbf{ }local hidden variable theory
(RLHVT) and quantum mechanics (QM) \cite{bell64,clauser69,bell04} to
experimental observations in the Aspect-Grangier-Roger experiment
\cite{aspect82}. First, the Bell inequality is initially expressed in terms of
$E_{\mathrm{RLHVT}}(\hat{a}$, $\hat{b})$-like coefficients with
$E_{\mathrm{RLHVT}}(\hat{a}$, $\hat{b})\overset{\text{def}}{=}\int\bar
{A}\left(  \hat{a}\text{, }\lambda\right)  \bar{B}(\hat{b}$, $\lambda
)\rho\left(  \lambda\right)  d\lambda$ in Eq. (\ref{scc2}). Second,
identifying $E_{_{\mathrm{QM}}}(\hat{a}$, $\hat{b})\overset{\text{def}%
}{=}\left\langle \psi|\left(  \vec{\sigma}\cdot\hat{a}\right)  \otimes
(\vec{\sigma}\cdot\hat{b})|\psi\right\rangle $ with $E_{\mathrm{RLHVT}}%
(\hat{a}$, $\hat{b})$, the Bell inequality can be recast in terms of
$E_{_{\mathrm{QM}}}(\hat{a}$, $\hat{b})$-like coefficients. Third, one shows
that for any entangled quantum state $\left\vert \psi\right\rangle $ with
$\left\langle \psi|\psi\right\rangle =1$, the Bell inequality in terms of
$E_{_{\mathrm{QM}}}(\hat{a}$, $\hat{b})$-like coefficients can be violated
\cite{gisin91}. Finally, it can be verified that quantum mechanical
predictions, unlike theoretical predictions arising from a realistic local
hidden variable theory, agree with experimental observations expressed in
terms of $E_{\mathrm{AGR}\text{-\textrm{exp}}}(\hat{a}$, $\hat{b})$, where
\cite{aspect82}%
\begin{equation}
E_{\mathrm{AGR}\text{-\textrm{exp}}}(\hat{a}\text{, }\hat{b}%
)\overset{\text{def}}{=}\frac{R_{++}(\hat{a}\text{, }\hat{b})+R_{--}(\hat
{a}\text{, }\hat{b})-R_{+-}(\hat{a}\text{, }\hat{b})-R_{-+}(\hat{a}\text{,
}\hat{b})}{R_{++}(\hat{a}\text{, }\hat{b})+R_{--}(\hat{a}\text{, }\hat
{b})+R_{+-}(\hat{a}\text{, }\hat{b})+R_{-+}(\hat{a}\text{, }\hat{b})}\text{.}
\label{expco}%
\end{equation}
These logical steps can be outlined as follows:

\begin{enumerate}
\item[{[i]}] Derive the Bell inequality in the Clause-Horne-Shimony-Holt
(CHSH) form within the formalism of a realistic local hidden variable theory
(RLHVT),%
\begin{equation}
\left\vert E_{\mathrm{RLHVT}}(\hat{a}\text{, }\hat{b})-E_{\mathrm{RLHVT}}%
(\hat{a}\text{, }\hat{b}^{\prime})\right\vert +\left\vert E_{\mathrm{RLHVT}%
}(\hat{a}^{\prime}\text{, }\hat{b})+E_{\mathrm{RLHVT}}(\hat{a}^{\prime}\text{,
}\hat{b}^{\prime})\right\vert \leq2\text{.}%
\end{equation}

\item[{[ii]}] Identify $P(\hat{a}$, $\hat{b})\overset{\text{def}%
}{=}E_{_{\mathrm{QM}}}(\hat{a}$, $\hat{b})$ with $E_{\mathrm{RLHVT}}(\hat{a}$,
$\hat{b})$. Note that $P(\hat{a}$, $\hat{b})$ and $P\left(  a\text{,
}b\right)  $ in the main text denote the same quantity.

\item[{[iii]}] Rewrite the Bell inequality in [i] by using the identification
in [ii],%
\begin{equation}
\left\vert P(\hat{a}\text{, }\hat{b})-(P\hat{a}\text{, }\hat{b}^{\prime
})\right\vert +\left\vert P(\hat{a}^{\prime}\text{, }\hat{b})+P(\hat
{a}^{\prime}\text{, }\hat{b}^{\prime})\right\vert \leq2\text{.} \label{bell2}%
\end{equation}

\item[{[iv]}] Show that there are quantum mechanical states $\left\vert
\psi\right\rangle $ (i.e., entangled states) for which the Bell inequality in
Eq. (\ref{bell2}) is violated. For instance,%
\begin{equation}
\left\vert P(\hat{a}\text{, }\hat{b})-P(\hat{a}\text{, }\hat{b}^{\prime
})\right\vert +\left\vert P(\hat{a}^{\prime}\text{, }\hat{b})+P(\hat
{a}^{\prime}\text{, }\hat{b}^{\prime})\right\vert >2\text{.}%
\end{equation}
In particular, for a given entangled state $\left\vert \psi\right\rangle $,
the strength of the violation depends on the sets of orientations $(\hat{a}$,
$\hat{b})$, $(\hat{a}$, $\hat{b}^{\prime})$, $(\hat{a}^{\prime}$, $\hat{b})$,
and $(\hat{a}^{\prime}$, $\hat{b}^{\prime})$.

\item[{[v]}] Confirm, by means of the Aspect-Grangier-Roger experiment
\cite{aspect82}, the violation of the Bell inequality in the CHSH form once
$E_{\mathrm{AGR}\text{-\textrm{exp}}}(\hat{a}$, $\hat{b})$ in Eq.
(\ref{expco}) is identified with $E_{_{\mathrm{QM}}}(\hat{a}$, $\hat{b})$.
\end{enumerate}

In Table III, we report the expressions of spin correlation coefficients in a
realistic local hidden variable theory (RLHVT), in quantum mechanics (QM) and,
finally, in the Aspect-Grangier-Roger experiment (AGR-exp).\begin{table}[t]
\centering
\begin{tabular}
[c]{c|c|c}\hline\hline
\textbf{Theory and experiment} & \textbf{Spin correlation coefficient symbol}
& \textbf{Spin correlation coefficient expression}\\\hline
Realistic local hidden variable theory & $E_{\mathrm{RLHVT}}\left(  \hat
{a}\text{, }\hat{b}\right)  $ & $\int\bar{A}\left(  \hat{a}\text{, }%
\lambda\right)  \bar{B}\left(  \hat{b}\text{, }\lambda\right)  \rho\left(
\lambda\right)  d\lambda$\\
Quantum mechanics & $E_{\mathrm{QM}}\left(  \hat{a}\text{, }\hat{b}\right)  $
& $\left\langle \psi|\left(  \vec{\sigma}\cdot\hat{a}\right)  \otimes\left(
\vec{\sigma}\cdot\hat{b}\right)  |\psi\right\rangle $\\
The Aspect-Grangier-Roger experiment & $E_{\mathrm{AGR}\text{-\textrm{exp}}%
}\left(  \hat{a}\text{, }\hat{b}\right)  $ & $\frac{R_{++}\left(  \hat
{a}\text{, }\hat{b}\right)  +R_{--}\left(  \hat{a}\text{, }\hat{b}\right)
-R_{+-}\left(  \hat{a}\text{, }\hat{b}\right)  -R_{-+}\left(  \hat{a}\text{,
}\hat{b}\right)  }{R_{++}\left(  \hat{a}\text{, }\hat{b}\right)
+R_{--}\left(  \hat{a}\text{, }\hat{b}\right)  +R_{+-}\left(  \hat{a}\text{,
}\hat{b}\right)  +R_{-+}\left(  \hat{a}\text{, }\hat{b}\right)  }$\\\hline
\end{tabular}
\caption{Expressions of spin correlation coefficients in a realistic local
hidden variable theory (RLHVT), in quantum mechanics (QM) and, finally, in the
Aspect-Grangier-Roger experiment (AGR-exp).}%
\end{table}

\section{Conclusions}

In this paper, we presented a revisitation of the violation of Bell's
inequality in the CHSH form with entangled quantum states. First, we discussed
the 1935 EPR paradox that emerges from putting the emphasis on Einstein's
locality and the absolute character of physical phenomena. Second, we
discussed Bell's 1971 derivation of the 1969 CHSH form of the original 1964
Bell inequality in the context of a RLHVT. Third, identifying the
quantum-mechanical spin correlation coefficient with the RLHVT one, we
followed Gisin's 1991 analysis to show that QM violates Bell's inequality when
systems are in entangled quantum states. For pedagogical purposes, we showed
how this violation depends both on the orientation of the polarizers and the
degree of entanglement of the quantum states (see Fig. $1$, for instance).
Fourth, we discussed the basics of the experimental verification of Bell's
inequality in an actual laboratory as presented in the original 1982 AGR) experiment.

In what follows, we present a summary of take home messages from our
discussion on this beautiful piece of physical science with a deep interplay
between theory (i.e., scientific imagination and theoretical creativity) and
experiments (i.e., cleverly designed experimental set-ups and high statistical accuracy):

\begin{enumerate}
\item[{[i]}] From the EPR paradox \cite{epr}, one might intuitively think that
the statistical predictions of quantum mechanics emerge from a hidden
underlying deterministic substructure that encodes the natural assumption of locality.

\item[{[ii]}] John Bell derived Bell's inequality within the framework of a
realistic local-hidden variable theory \cite{bell64,bell04}.

\item[{[iii]}] Quantum mechanics is a non-local theory and violates Bell's
inequality \cite{bell64,bell04}.

\item[{[iv]}] Natural phenomena occur in a laboratory, not in a Hilbert space
\cite{peres2004}. The experiments with entangled photons by Aspect and
collaborators \cite{aspect82,aspect82B}, Clauser and collaborators
\cite{clauser69,clauser72} and, finally, Zeilinger and collaborators
\cite{pan98} experimentally establish the violation of Bell's inequalities and
give evidence of one of the most striking aspects of quantum physics (i.e.,
quantum entanglement).

\item[{[v]}] Entanglement is the key concept needed to understand the EPR
paradox and the violation of Bell's inequality by quantum mechanics
\cite{pan98,zeilinger98,brukner01}.

\item[{[vi]}] Unlike what one might intuitively believe, quantum entanglement
requires neither the entangled particles to come from a common source nor to
have interacted in the past. This fact was experimentally verified by
Zeilinger and collaborators in Ref. \cite{pan98}.

\item[{[vii]}] As pointed out by Zeiliger in Ref. \cite{zeilinger98},
information about quantum systems is more important than any possible
so-called \textquotedblleft\emph{real}\textquotedblright\ property these
systems might possess. Finally, there appears to exist no quantum world.
Instead, as pointed out by Bohr in Ref. \cite{bohr35}, there is only an
abstract quantum-mechanical description of natural phenomena.
\end{enumerate}

\smallskip

This set of take home messages concludes our revisitation of the violation of
Bell's inequality in the Clauser-Horne-Shimony-Holt form with entangled
quantum states. Interestingly, our idea of this revisitation started at the
end of 2021 and, with our gladness, it is a very nice coincidence that the
Nobel prize in Physics was assigned in 2022 to Aspect, Clauser, and Zeilinger
\cite{epjd22}. Clearly, we express our congratulations to these three
physicists along with their research Groups. Finally, we wholeheartedly
believe this whole chain of events- EPR, Bohr, Bell, entanglement, interplay
between theory and experiment- is the essence of physics at its best. We are
aware that our effort here is in no way representative of this wonderful
scientific endeavor. It would have been too big a task for us to accomplish in
the present scientific effort. However, being this paper written by two
theorists, a young graduate student, and an experimental physicist, we hope it
will help young students and future generations of researchers appreciate (via
a description of a sequence of actual scientific events) how important are the
processes of \emph{scientific imagination} and \emph{experimental
verification} in modern physics.

\begin{acknowledgments}
C. Cafaro thanks Ariel Caticha, Domenico Felice, Nicholas Gisin, Yanjun He,
and Elena R. Loubenets for valuable comments and helpful electronic mail
exchanges throughout the last years that helped shaping this paper in its
current form.
\end{acknowledgments}

\end{document}